\documentclass[aps,twocolumn]{revtex4}
\usepackage [dvips]{graphicx}
\usepackage{multirow,amsmath,ulem}
\usepackage{color}
\usepackage{rotating}

\begin{document}

\title{Equation of state and thickness of the inner crust of neutron stars}

\author{Fabrizio Grill}
\affiliation{Centro de F\'{i}sica Computacional, Department of Physics, University of Coimbra, P-3004-516 Coimbra, Portugal}
\author{Sidney S. Avancini}
\affiliation{Depto de F\'{\i}sica - CFM - Universidade Federal de Santa Catarina  Florian\'opolis - SC - CP. 476 - CEP 88.040 - 900 - Brazil}
\author{Helena Pais}
\affiliation{Centro de F\'{i}sica Computacional, Department of Physics, University of Coimbra, P-3004-516 Coimbra, Portugal}
\author{Constan\c{c}a Provid\^encia}
\affiliation{Centro de F\'{i}sica Computacional, Department of Physics, University of Coimbra, P-3004-516 Coimbra, Portugal}
\author{Isaac Vida\~na}
\affiliation{Centro de F\'{i}sica Computacional, Department of Physics, University of Coimbra, P-3004-516 Coimbra, Portugal}

\begin{abstract}
The cell structure of $\beta$-stable clusters in the inner crust of cold and  warm
neutron stars is studied within the Thomas--Fermi
approach using relativistic mean field nuclear models. The relative size of the
inner crust and the pasta phase of neutron stars is
calculated, and the effect of the symmetry energy slope parameter, $L$, on the profile of
the neutron star crust is discussed. It is shown that while the
size of the  total crust is {mainly} determined by the incompressibility modulus, the relative
size of the inner crust depends on $L$. It is found that the inner crust represents a larger fraction of 
the total crust for smaller values of $L$. Finally, it is shown that at finite temperature the pasta phase in 
$\beta$-equilibrium matter essentially melts above $5-6$ MeV, and that the onset
density of the rodlike and slablike structures does not depend on the temperature.
\end{abstract}
\maketitle
\vspace{0.50cm}
PACS number(s): {21.65.Cd, 24.10.Jv, 26.60.-c, 97.60.Jd}
\vspace{0.50cm}

\section{Introduction}

Nowadays, it is generally  accepted the existence  of the so-called
{\it pasta phase} \cite{pas1,pas2,pas3,pas4,Haensel,Maruyama, pasta1,pasta2,pasta3,pais12}
in the inner crust of a neutron star close to the  crust-core transition.
Constituted by  several types of complex structures such as  {\it e.g.,} rods and slabs, the pasta phase is
a frustrated system formed as a result of the competition between the strong and the electromagnetic interactions. 

The effect of the density dependence of  the symmetry energy  on the pasta
phase has been discussed in several works  \cite{oyamatsu07,grill12,pasta4,providencia13}.
In particular, it has been shown that for very large values of the symmetry energy slope parameter, 
$L=3n_0(\partial E_{sym}(n)/\partial n )_{n_0}$,
non-spherical structures ({\it e.g.,} rodlike or slablike)  are not expected in
$\beta$-equilibrium matter, and that the number of
nucleons in the clusters as well as the cluster proton fraction and the size of the 
Wigner--Seitz (WS) cell are sensitive to this quantity. It has been also discussed
that $L$ may have quite dramatic effects on the cell structure if its value is very large 
or small \cite{pasta3,grill12,pasta4}.

In the present work  the effect of $L$ on the size of the inner crust will be discussed 
within a Thomas--Fermi (TF) formalism  in
the WS approximation in the framework of relativistic mean-field (RMF)
nuclear models \cite{Maruyama,pasta1,pasta2,pasta3}. 
The Tolman--Oppenheimer--Volkov equations (TOV) \cite{tov} will be integrated
and the size of the inner crust as well as the localization of the pasta structures identified.
In particular, it
will be  shown that smaller values of $L$ favor a wider slab phase and a larger
relative size of the inner crust with respect to the total crust, and a steeper crust profile.

We will also study the effect of temperature on 
the size of the inner crust.
It will be shown
that pasta clusters in $\beta$-equilibrium  will completely melt for temperatures above 5-6 MeV. 
These results agree partially with the
predictions obtained within a dynamical spinodal (DS) approach \cite{cp2006,brito2006,pais}. A similar calculation was done with Skyrme forces in Ref.\ \cite{ducoin08}.  The melting of the pasta phase 
was previously studied in \cite{melt1,melt2}, were the effect of
thermal fluctuations was taken into account.  It is expected that if
thermal fluctuations are larger than the radius of the WS
cell, the  WS approximation breaks {down}.  
However, thermal fluctuations will not be considered in the present calculation and, therefore, 
the reader must interpret our results just as upper limits.
The existence of non-homogeneous matter will affect the evolution of a supernova or proto-neutron star matter, in particular the diffusion of neutrinos out of the star \cite{pas3}.
Some preliminary results of the present study have been published in Ref.\ \cite{providencia13}.

The paper is organized as follows. The formalism is briefly reviewed in section\ \ref{sec2}. Section \ref{sec3} is  devoted to the presentation and discussion 
of the results while the main conclusions are given in section \ref{sec4}.

\section{Formalism}
\label{sec2}

To describe the inner crust in this work we apply the self-consistent
TF formalism presented in Refs.\ \cite{pasta1,pasta2,pasta3}. We use
relativistic mean field nuclear models 
with constant couplings and non-linear terms~\cite{bb}, and with density dependent
couplings~\cite{tw}. Within the first class of models, that we will
designate by Non Linear Walecka Models, we consider the 
following ones: NL3~\cite{nl3} with non linear $\sigma$ terms,
TM1 \cite{tm1} with non linear $\sigma$ and $\omega$ terms,
NL3$_{\omega\rho}$ including also non-linear $\omega\rho$ terms which allow the
modulation of the density dependence of the symmetry
energy~\cite{hor01}, FSU~\cite{fsu} and IU-FSU~\cite{iufsu} with
non-linear $\sigma$, $\omega$ and $\omega\rho$ terms. 
Within the second class of models with density dependent couplings
we consider the models DD-ME2~\cite{ddme2} and DD-ME$\delta$~\cite{roca2011}.
The latter, among the parametrizations considered, is the only one
including the $\delta$ meson.

All the equations that allow the performance of the TF
calculation  are derived from the Lagrangian density 
\begin{equation}
\mathcal{L}=\sum_{i=n,p}\mathcal{L}_{i}+\mathcal{L}_e\mathcal{\,+L}_{{\sigma }}%
\mathcal{+L}_{{\omega }}\mathcal{+L}_{{\rho
}}\mathcal{+L}_{{\delta}}\mathcal{+L}_{{\gamma }}
\mathcal{+L}_{{nl}}, \label{lagdelta}
\end{equation}
where the nucleon Lagrangian reads
\begin{equation}
\mathcal{L}_{i}=\bar{\psi}_{i}\left[ \gamma _{\mu }iD^{\mu }-M^{*}\right]
\psi _{i}  \label{lagnucl},
\end{equation}
with
\begin{eqnarray}
iD^{\mu } &=&i\partial ^{\mu }-\Gamma_{\omega}\omega^{\mu }-\frac{\Gamma_{\rho }}{2}{\boldsymbol{\tau}}%
\cdot \boldsymbol{\rho}^{\mu } - e \frac{1+\tau_3}{2}A^{\mu}, \label{Dmu} \\
M^{*} &=&M-\Gamma_{\sigma}\sigma-\Gamma_{\delta
}{\boldsymbol{\tau}}\cdot \boldsymbol{\delta}, \label{Mstar}
\end{eqnarray}
and the electron Lagrangian is given by
\begin{equation}
\mathcal{L}_e=\bar \psi_e\left[\gamma_\mu\left(i\partial^{\mu} + e A^{\mu}\right)
-m_e\right]\psi_e.
\label{lage}
\end{equation}
The meson and electromagnetic Lagrangian densities are
\begin{eqnarray*}
\mathcal{L}_{{\sigma }} &=&\frac{1}{2}\left( \partial _{\mu }\sigma \partial %
^{\mu }\sigma -m_{\sigma}^{2}\sigma ^{2}\right)  \\
\mathcal{L}_{{\omega }} &=&\frac{1}{2} \left(-\frac{1}{2} \Omega _{\mu \nu }
\Omega ^{\mu \nu }+ m_{\omega}^{2}\omega_{\mu }\omega^{\mu } \right) \\
\mathcal{L}_{{\rho }} &=&\frac{1}{2} \left(-\frac{1}{2}
\mathbf{R}_{\mu \nu }\cdot \mathbf{R}^{\mu
\nu }+ m_{\rho }^{2}\boldsymbol{\rho}_{\mu }\cdot \boldsymbol{\rho}^{\mu } \right)\\
\mathcal{L}_{ {\delta }} &=&\frac{1}{2}(\partial _{\mu }\boldsymbol{\delta}%
\partial ^{\mu }\boldsymbol{\delta}-m_{\delta }^{2}{\boldsymbol{\delta}}^{2})\\
\mathcal{L}_{{\gamma }} &=&-\frac{1}{4}F _{\mu \nu }F^{\mu
  \nu }\\
\mathcal{L}_{{nl}} &=&-\frac{1}{3!}\kappa \sigma ^{3}-\frac{1}{4!}%
\lambda \sigma ^{4}+\frac{1}{4!}\xi \Gamma_{\omega}^{4}(\omega_{\mu}\omega^{\mu })^{2} \\
&+&\Lambda_\omega \Gamma_\omega^2 \Gamma_\rho^2 \omega_{\nu
}\omega^{\nu } \boldsymbol{\rho}_{\mu }\cdot
\boldsymbol{\rho}^{\mu }
\end{eqnarray*}
where $\Omega _{\mu \nu }=\partial _{\mu }\omega_{\nu }-\partial
_{\nu }\omega_{\mu }$, $\mathbf{R}_{\mu \nu }=\partial _{\mu
}\boldsymbol{\rho}_{\nu }-\partial _{\nu }\boldsymbol{\rho}_{\mu
}-\Gamma_{\rho }(\boldsymbol{\rho}_{\mu }\times
\boldsymbol{\rho}_{\nu })$ and $F_{\mu \nu }=\partial _{\mu
}A_{\nu }-\partial _{\nu }A_{\mu }$. The four coupling parameters
$\Gamma_\sigma$, $\Gamma_\omega$, $\Gamma_{\rho}$ and
$\Gamma_{\delta}$ of the mesons to the nucleons are density
dependent in the relativistic density dependent models considered,
namely, DD-ME2~\cite{ddme2} and DD-ME$\delta$~\cite{roca2011}. The
non-linear term $\mathcal{L}_{{nl}} $ is absent in these models.
In all the other models, NL3~\cite{nl3}, TM1 \cite{tm1},
NL3$_{\omega\rho}$~\cite{hor01}, FSU~\cite{fsu} and
IU-FSU~\cite{iufsu}, the couplings are constant and at least some
of the non-linear terms of $\mathcal{L}_{nl}$ are included. In the
above Lagrangian density $\boldsymbol {\tau}$ is the isospin
operator. For reference, we give in Table \ref{TabSat} the main
properties of the above models at saturation density. We will discuss how sensitive is the structure of the
non-homogeneous inner-crust of a neutron star to the properties of
the Equation of State (EoS).

In the TF approximation of non-uniform $npe$ matter, fields are assumed to vary
slowly so that baryons can be treated as moving in locally constant fields at 
each point \cite{Maruyama, pasta1}. The calculation starts from the grand canonical potential 
density,
\begin{eqnarray}
\omega&=&\omega(\{f_{i+}\},\{f_{i-}\} ,\sigma,\omega_0,\rho_{30},\delta_0) \nonumber \\
&=&{\mathcal E}-T{\mathcal S}-\sum_{i=n,p,e}\mu_in_i
\label{gc}
\end{eqnarray}
where $\{f_{i+}\} (\{f_{i-}\})$ with $i=n,p,e$ stands for the neutron, proton and electron positive (negative)
energy distributions, and 
${\mathcal S}$ and ${\mathcal E}$ are the total entropy and energy densities, respectively 
\cite{pasta3}. The equations of motion for the meson fields (see {\it e.g.,} Ref.\ \cite{pasta1} for details) follow from the 
variational conditions
\begin{equation} 
\frac{\delta\Omega}{\delta\sigma(\bf r)}=\frac{\delta\Omega}{\delta\omega_0(\bf r)}=
\frac{\delta\Omega}{\delta\rho_{30}(\bf r)}=\frac{\delta\Omega}{\delta\delta_0(\bf r)}=0 \ ,
\label{vc}
\end{equation}
where $\Omega=\int d^3r\,\omega$.
The numerical algorithm for the description of the neutral $npe$ matter at finite temperature is a generalization
of the $T=0$ case which is discussed in detail in Refs. \cite{pasta1,pasta3}. The Poisson equation is always solved
by using the appropriate Green's function according to the spatial dimension of interest, and the Klein--Gordon
equations are solved by expanding the meson fields in a harmonic oscillator basis with one, two or three dimensions
based on the method presented in Refs.\ \cite{pasta1, pasta3}. The interested reader is referred to these works
for details of the calculation.

\begin{table}
  \centering
  \begin{tabular}{l|ccccccc}
    \hline\hline
model & $n_0$           & $E_0$ &     $K_0$ &    $Q_0$ & $E_{sym}$ &  $L$ \\
      & $(\mbox{fm}^{-3})$ & (MeV) &    (MeV)  &    (MeV) &     (MeV) & (MeV) \\
\hline
    NL3                & $0.148$ & $-16.24$ & $270.7$ & 203 & $37.3$ & $118.3$ \\
    TM1                & $0.145$ & $-16.26$ & $280.4$ &-295 & $36.8$ & $110.6$ \\
    FSU                & $0.148$ & $-16.30$ & $230.0$ &-523 & $32.6$ & $60.5$  \\
    NL3$_{\omega\rho}$ & $0.148$ & $-16.30$ & $272.0$ & 203 & $31.7$ & $55.2$  \\
    DD-ME$\delta$      & $0.152$ & $-16.12$ & $219.1$ &-741 & $32.4$ & $52.9$  \\
    DD-ME2             & $0.152$ & $-16.14$ & $250.8$ & 478 & $32.3$ & $51.4$  \\
    IU-FSU             & $0.155$ & $-16.40$ & $231.2$ &-288 & $31.3$ & $47.2$
\\
    \hline\hline
  \end{tabular}
  \caption{Nuclear matter properties at saturation: density ($n_0$),
    energy ($E_0$), incompressibility ($K_0$), skewness ($Q_0$), 
    symmetry energy ($E_{sym}$), and symmetry energy slope parameter ($L$).}
\label{TabSat}
\end{table}

\section{Results}
\label{sec3}

In this section we present and discuss the results obtained for the different models 
considered. The section is divided in three parts. In the first one our attention is
focused on the sensitivity of the thickness and structure of the inner crust to
the properties of the EoS. In the second one we discuss the effect of $L$ on
the density profile of the crust.  Finally, in the last part we analyze the effect of finite 
temperature on the crust and the pasta phase. The discussion is done for neutron stars 
with masses $M=1, 1.44$ and $1.6 M_\odot$.  The first two values have been chosen 
as representative masses since $1 M_\odot$ is smaller than the smallest neutron star mass 
detected until now and $1.44 M_\odot$ is the mass of the famous Hulse--Taylor pulsar. The value
of $1.6 M_\odot$ is chosen to be slightly smaller than the maximum mass predicted by the FSU
model. Results for the maximum neutron star mass configuration have also been obtained
for all the models.
\begin{figure}[tb]
\begin{tabular}{c}
  \includegraphics[width=0.8\linewidth]{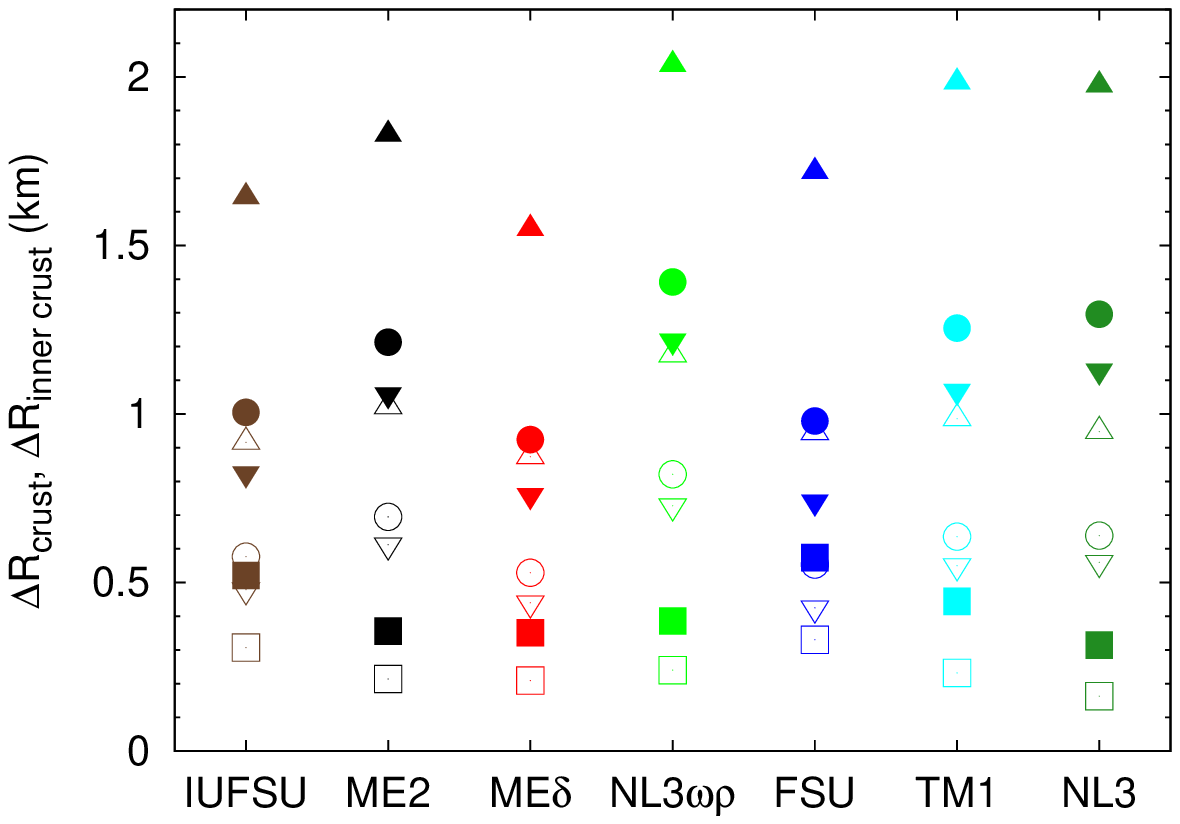}\\
  \includegraphics[width=0.8\linewidth]{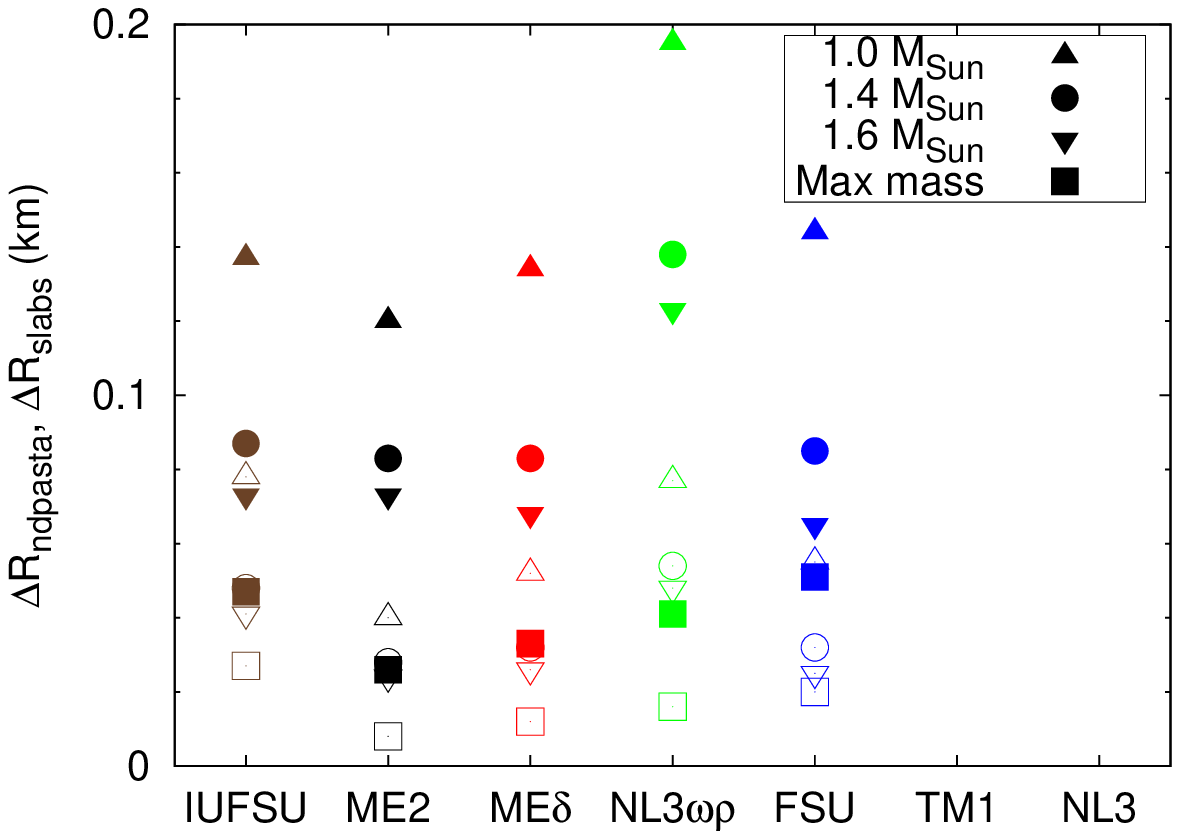}
  \end{tabular}
  \caption{(Color online) Top panel: Crust (full symbols) and inner crust (empty
    symbols) thickness. Bottom panel : Thickness of
    the non-droplet pasta  (ndpasta)  (full symbols) and the slab (empty
    symbols) phases. The symbol shape identifies the star mass: 
    1.0 $M_\odot$ (upward triangle), 1.44 $M_\odot$ (circle), 1.6 $M_\odot$ (downward triangle)
    and maximum mass (square).}
\label{crust1}
\end{figure}

\begin{figure}[t]
\begin{tabular}{c}
  \includegraphics[width=0.7\linewidth]{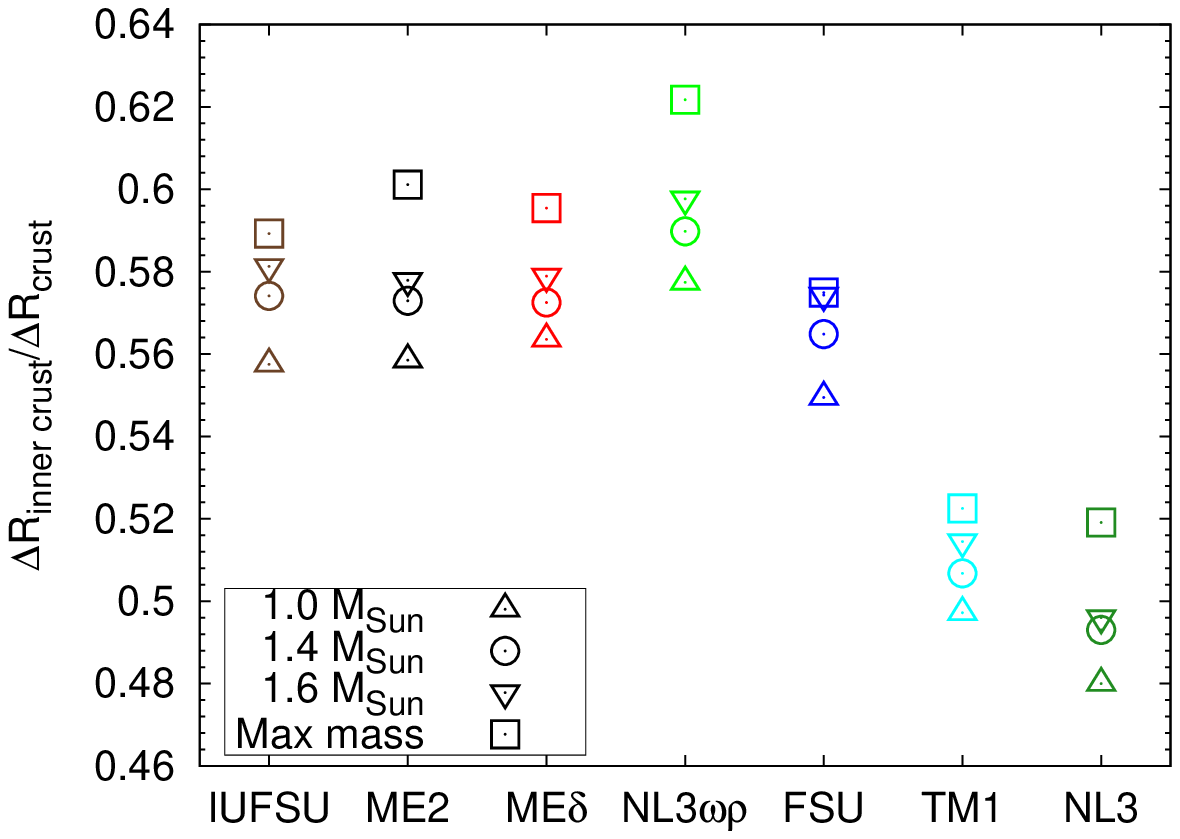}\\
  \includegraphics[width=0.7\linewidth]{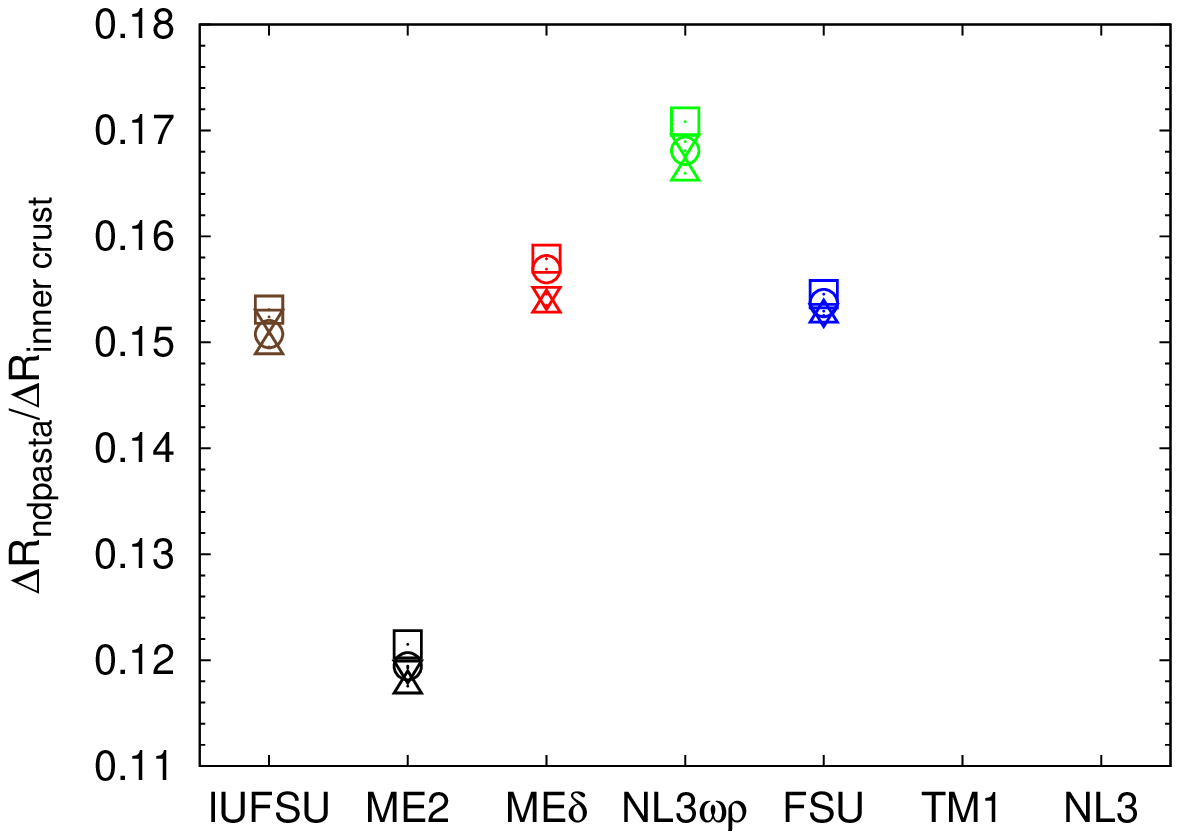}\\
  \includegraphics[width=0.7\linewidth]{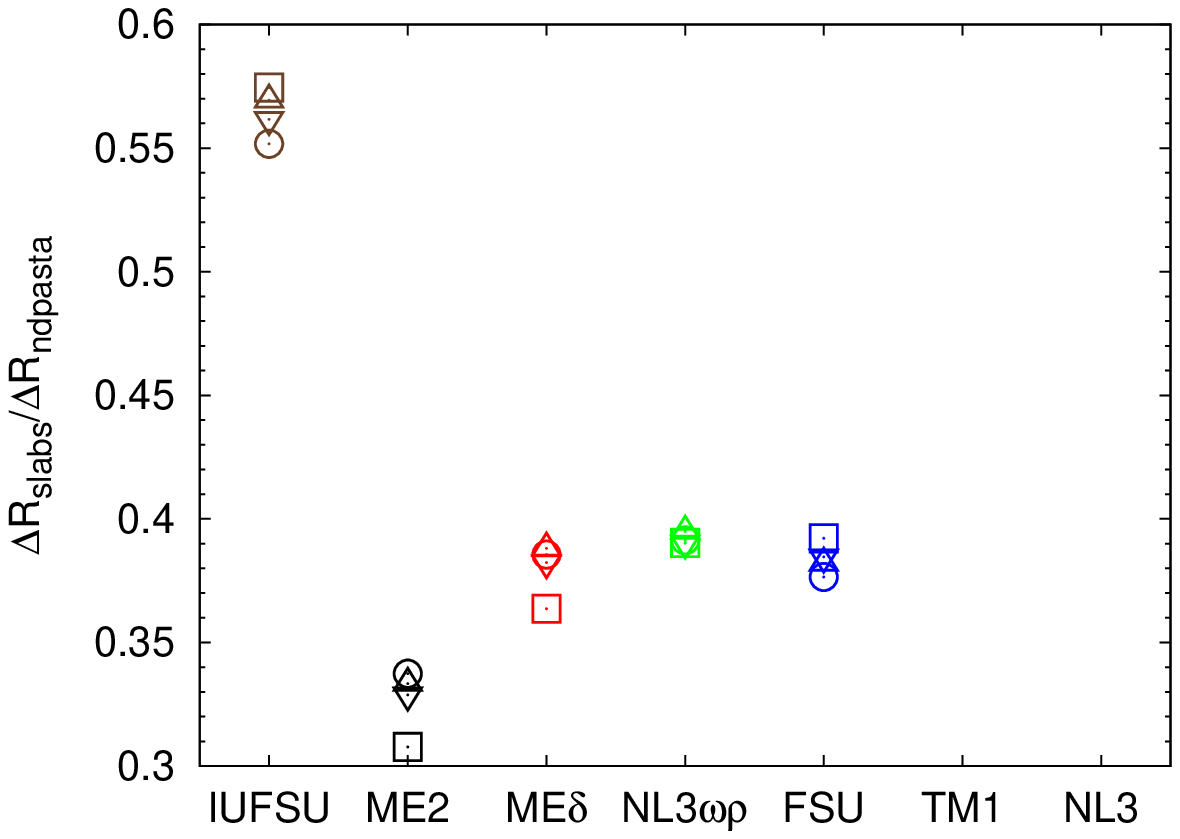}
\end{tabular}
\caption{(Color online) Top panel: Fraction of the crust
  occupied by the the inner crust. Middle panel: Fraction of the inner crust occupied
  by the non-droplet pasta (ndpasta)  phase. Bottom panel: Fraction of the
  non-droplet pasta phase occupied by the slab
  phase. The symbol shape has the same meaning as in Fig. \ref{crust1}.}
\label{crust2}
\end{figure}

\subsection{Thickness and structure of the inner crust}

We present here results for the pasta phase of $\beta$-equilibrium non-homogeneous matter obtained within
our TF calculation at $T=0$. As done by many authors, we assume that for some given conditions (temperature,
density, proton fraction or chemical equilibrium) only a single geometry will be the physical one, {\it i.e.,} the one
with smaller free energy  in comparison with homogeneous
$\beta$-stable matter.  At least five different geometries could in principle 
occur: droplets, rods, slabs, tubes and bubbles.  However, due to the $\beta$-equilibrium the proton fraction is 
very small and only three of them are found energetically favorable: droplets, rods and slabs. These 
structures form a regular lattice that we study in the WS approximation. The TF approach is semiclassical 
and does not include shell effects. Nevertheless,  it has been recently shown \cite{grill12} that the main 
properties of the WS cells obtained within a TF calculation agree with  Hartree--Fock (HF) \cite{NV} and
Hartree--Fock--Boguliobov (HFB) \cite{grill} calculations which allow the inclusion of shell effects. For a 
comparison of HFB and TF results the interested reader is referred to Ref.\  \cite{grill12}.  

The complete stellar matter EoS is built by properly joining the inner crust part with the outer crust and the core ones.
In this work, we assume that the core of the star is made only of nucleons, electrons and muons, and its EoS is obtained
also in RMF approach imposing both $\beta$-equilibrium and charge neutrality. For the outer crust we consider the
well known Baym--Pethick--Sutherland (BPS) EoS \cite{bps}. The TOV equations are then solved to determine the 
density profile of the neutron stars with the masses $M=1,1.44, 1.6 M_\odot$ and $M_{max}$ mentioned before. 

In Table \ref{profile} we show for the different models some of the features of the inner crust structure. All the models 
have a slab and a rod phase which together define the thickness of the non-droplet pasta, except for the NL3 and TM1 models. 
For these two models the inner crust is only formed by droplets in a neutron gas background. This is, as shown in 
Ref.\ \cite{oyamatsu07}, due to the high value of $L$  for these two models,  118 MeV  and 110 MeV, respectively.

{In order to help} the discussion, the results of Table \ref{profile} are also 
plotted, in Figs.\ \ref{crust1} and  \ref{crust2}. In Fig.\  \ref{crust1},  the thickness 
of the {total} crust (full symbols) and inner crust (empty symbols) are shown in the top panel, and  the thickness 
of the total non-droplet pasta phase (full symbols) and the slab phase (empty symbols) are plotted in the bottom panel.
In Fig.\  \ref{crust2}  we plot the ratios of these quantities: the fraction of the inner crust with respect to the total crust (top
panel), the fraction of the non-droplet pasta extension with respect to the total inner crust (middle panel) and the fraction of slab phase
size with respect to the total non-droplet phase (bottom panel). The different models are ordered according to the magnitude of the 
slope $L$,  which {increases} from left to right.

\begin{figure}
 \begin{tabular}{c}
  \includegraphics[width=0.8\linewidth]{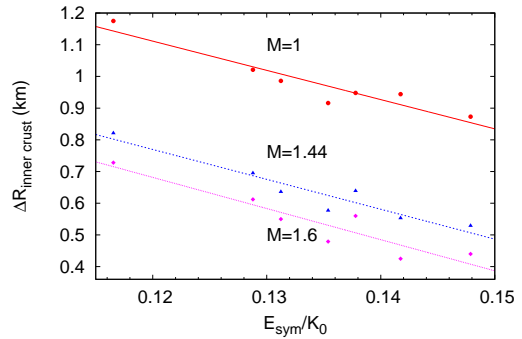}\\
    \end{tabular}
  \caption{(Color online) Correlation between the inner crust thickness and
    the  ratio $E_{sym}/K_0$ for star with $M=1,\, 1.44, \, 1.6\, M_\odot$.
  The slope of the three straight lines is similar $\sim -9.5 \pm 15\%$
km.}
\label{jkcrust}
\end{figure}

From Fig.\ \ref{crust1} we see that no clear trend is found in the thickness of the different parts as a function of $L$.
Instead, {we have found (see Fig.\ \ref{jkcrust}) that}  the size of the total crust is mainly defined by the incompressibility of the EoS (cfr
Table~\ref{TabSat}). In particular, the models NL3, NL3$\omega\rho$, TM1, followed by DDME2, have
the largest crusts and the largest incompressibilities.  In general, a systematic  behaviour is observed with the mass 
of the star:  the larger the mass, the thinner the crust and its sub-layers. 
{ For instance, in all cases} for $M=1 \, M_\odot$ the crust thickness lies between 1.5 and 2 km, 
while for the maximum mass configurations it is mainly below 0.5 km.

The top panel of Fig. \ref{crust2} shows that the inner crust occupies $\gtrsim$ 55\% of the total crust size, 
except for NL3 and TM1, for which it reduces to $\sim$ 50\% of it. Once more, a systematic dependence upon the star mass is found: 
the fraction of the crust occupied by the inner crust increases with the star mass and there is a
difference of $\sim + 2\%$ between a star with  {1 and 1.6 $M_\odot$.}

Next, we come to the size of the non-droplet pasta phase, the phases
that correspond to a frustrated system.  In a 1 $M_\odot$ star the non-droplet pasta size is smaller than 
200 m (Fig. \ref{crust1} bottom). Its relative size  is slightly smaller in less massive stars (Fig. \ref{crust2} middle). 
Within the non-droplet pasta phase, we find in our results slabs and rods phases. The slab fraction corresponds to 
$\sim35-40\%$ of the total non-droplet pasta phase for all the models, apart from IU-FSU, where it is almost $60\%$ 
and DDME2  with $\sim30-35\%$  (Fig. \ref{crust2} bottom). The different behavior of IU-FSU is mainly due to the small 
proton fraction in the cluster. IU-FSU has a small value of the symmetry energy slope at subsaturation densities, which affects 
the surface tension giving quite a high surface tension, see \cite{grill12}, and
preventing the neutron drip. A small proton fraction in the cluster favors the slab geometry with respect to the
rod geometry because the surface energy decreases with the proton fraction. On the contrary, a smaller surface tension favors the
neutron drip.  Clusters are more isospin symmetric and the droplet geometry is favored. This could explain  the behavior of DDME2 with 
the smallest fraction of the non-droplet pasta phase.  In Ref.\ \cite{grill12} it is shown that the DDME2 and NL3 models have the smallest 
surface energies for nuclear symmetric matter.

In Ref.\ \cite{newton2011} the effect of the nuclear pasta on the crustal shear phenomena was studied. In particular, two limits were 
considered, namely the pasta as an elastic solid and as a liquid. In
the first case the shear modulus was calculated at  the  
crust-core transition while in the second case it was done at the transition from the droplet phase to the non-droplet pasta phase. For
models with no non-droplet pasta phase, such as NL3 and TM1, there is no difference between these two pictures. However, models with a
symmetry energy slope $L$ below 80 MeV, have a non-droplet pasta phase, and the ratio shear modulus to pressure {can be} as high 
as two times larger if the pasta is considered an elastic solid and  $L=40$ MeV. An intermediate picture would consider the rod phase as an elastic 
solid and the slab phase as a liquid phase.

Comparing our results with the ones discussed in \cite{newton2011} a couple of comments are in order. First, the incompressibility of the EoS 
seems to have an important influence on the total crust, so that DDME2 with $L=51$ MeV has a larger crust than FSU with $L=60$
MeV. Second, except for NL3$\omega\rho$ with a large $K_0$, all the other models predict the non-droplet pasta phase of the 1.44 $M_\odot$ star 
with a thickness of $\sim 80$ m, similar to the one calculated in \cite{newton2011}.

As mentioned before both the incompressibility and the density
dependence of the symmetry energy affect the size of the inner crust. We have obtained 
a possible correlation between the ratio $E_{sym}/K_0$ and the inner crust thickness. This correlation is shown in Fig. \ref{jkcrust}. This should be 
further investigated and confirmed with a larger set of models. The correlation is worse for the $1.6\, M_\odot$ star, probably
because this mass lies closer to the maximum mass configuration. The slope obtained for the three star masses analysed is $\sim -9.5 \pm 15\%\,$ 
Km.

\begin{table}[tbh]
  \centering
  \begin{tabular}{cccccccc}
\hline
\hline
   $M$ &  $\rho_c$  & ${\mathcal E}_c$ &   $R_{h-s}$  &   $R_{s-r}$ &    $R_{r-d}$  &  $R_{d-BPS}$ &   $R$\\
 ($M_\odot$) & (fm$^{-3}$) & (fm$^{-4}$) & (km)  & (km)  & (km)  &  (km) &(km) \\
\hline \multicolumn{8}{c}{NL3}\\
 $1.00$ & $ 0.226$ & $ 1.121$ & $12.568^*$ &     -    &     -    & $13.516$ & $14.543$ \\
 $1.44$ & $ 0.276$ & $ 1.390$ & $13.341^*$ &     -    &     -    & $13.980$ & $14.637$ \\
 $1.60$ & $ 0.293$ & $ 1.489$ & $13.534^*$ &     -    &     -    & $14.094$ & $14.663$ \\
 $2.78$ & $ 0.669$ & $ 4.415$ & $12.978^*$ &     -    &     -    & $13.141$ & $13.292$ \\
\multicolumn{8}{c}{TM1}\\
 $1.00$ & $ 0.243$ & $ 1.208$ & $12.425^*$ &     -    &     -    & $13.411$ & $14.408$ \\
 $1.44$ & $ 0.328$ & $ 1.674$ & $13.012^*$ &     -    &     -    & $13.648$ & $14.267$ \\
 $1.60$ & $ 0.366$ & $ 1.893$ & $13.089^*$ &     -    &     -    & $13.639$ & $14.158$ \\
 $2.18$ & $ 0.851$ & $ 5.345$ & $11.937^*$ &     -    &     -    & $12.169$ & $12.381$ \\
\multicolumn{8}{c}{FSU}\\
 $1.00$ & $ 0.356$ & $ 1.814$ & $11.088$ & $11.143$ & $11.232$ & $12.032$ & $12.806$ \\
 $1.44$ & $ 0.564$ & $ 3.073$ & $11.251$ & $11.283$ & $11.336$ & $11.804$ & $12.230$ \\
 $1.60$ & $ 0.757$ & $ 4.394$ & $10.921$ & $10.946$ & $10.986$ & $11.346$ & $11.661$ \\
 $1.66$ & $ 1.105$ & $ 7.100$ & $10.270$ & $10.290$ & $10.321$ & $10.600$ & $10.844$ \\
\multicolumn{8}{c}{NL3$_{\omega\rho}$}\\
 $1.00$ & $ 0.256$ & $ 1.277$ & $11.397$ & $11.474$ & $11.592$ & $12.572$ & $13.432$ \\
 $1.44$ & $ 0.298$ & $ 1.518$ & $12.365$ & $12.419$ & $12.503$ & $13.186$ & $13.757$ \\
 $1.60$ & $ 0.314$ & $ 1.616$ & $12.624$ & $12.672$ & $12.747$ & $13.352$ & $13.842$ \\
 $2.68$ & $ 0.680$ & $ 4.640$ & $12.479$ & $12.495$ & $12.520$ & $12.719$ & $12.865$ \\
\multicolumn{8}{c}{DD-ME$\delta$}\\
 $1.00$ & $ 0.405$ & $ 2.040$ & $10.550$ & $10.602$ & $10.684$ & $11.423$ & $12.099$ \\
 $1.44$ & $ 0.552$ & $ 2.890$ & $10.913$ & $10.945$ & $10.996$ & $11.442$ & $11.837$ \\
 $1.60$ & $ 0.627$ & $ 3.360$ & $10.901$ & $10.927$ & $10.969$ & $11.341$ & $11.661$ \\
 $1.96$ & $ 1.214$ & $ 7.938$ & $ 9.843$ & $ 9.855$ & $ 9.876$ & $10.052$ & $10.194$ \\
\multicolumn{8}{c}{DD-ME2}\\
 $1.00$ & $ 0.289$ & $ 1.435$ & $11.177$ & $11.217$ & $11.297$ & $12.198$ & $13.005$ \\
 $1.44$ & $ 0.347$ & $ 1.755$ & $12.021$ & $12.049$ & $12.104$ & $12.716$ & $13.234$ \\
 $1.60$ & $ 0.371$ & $ 1.895$ & $12.232$ & $12.256$ & $12.305$ & $12.844$ & $13.291$ \\
 $2.49$ & $ 0.817$ & $ 5.345$ & $11.717$ & $11.725$ & $11.743$ & $11.931$ & $12.073$ \\
\multicolumn{8}{c}{IU-FSU}\\
 $1.00$ & $ 0.342$ & $ 1.780$ & $10.820$ & $10.898$ & $10.957$ & $11.736$ & $12.463$ \\
 $1.44$ & $ 0.474$ & $ 2.607$ & $11.292$ & $11.340$ & $11.379$ & $11.869$ & $12.297$ \\
 $1.60$ & $ 0.558$ & $ 3.182$ & $11.265$ & $11.306$ & $11.338$ & $11.744$ & $12.089$ \\
 $1.81$ & $ 0.987$ & $ 6.662$ & $10.482$ & $10.509$ & $10.529$ & $10.789$ & $11.003$ \\
    \hline\hline
  \end{tabular}
  \caption{Central density ($\rho_c$) and energy density (${\mathcal E}_c$), distance to the center of the star at the
  phase transitions: homogeneous matter--slab phase ($ R_{h-s}$),
slab phase--rod phase ($R_{s-r}$), rod phase--droplet phase  ($
R_{r-d}$), droplet phase--outer crust ($R_{d-BPS}$) and radius of
a 1.0 $M_\odot$, 1.44 $M_\odot$, 1.6 $M_\odot$ and maximum mass
neutron star for the all the RMF
models considered. For NL3 and TM1 there are neither slabs nor rods: the values 
marked with an asterisk correspond to the homogeneous matter--droplet phase
transition.}\label{profile}
\end{table}

\begin{figure*}[t]
  \begin{tabular}{ccc}
  \includegraphics[angle=0,width=0.33\linewidth]{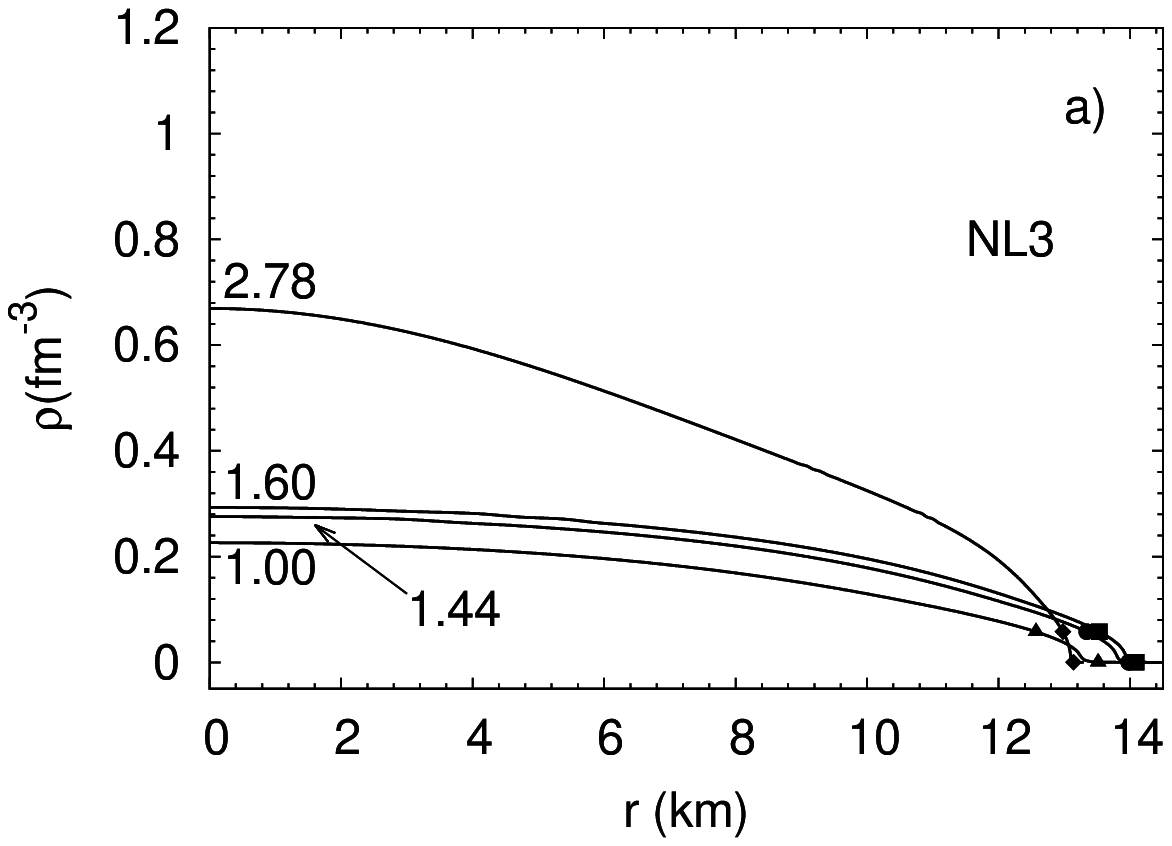}
  \includegraphics[angle=0,width=0.33\linewidth]{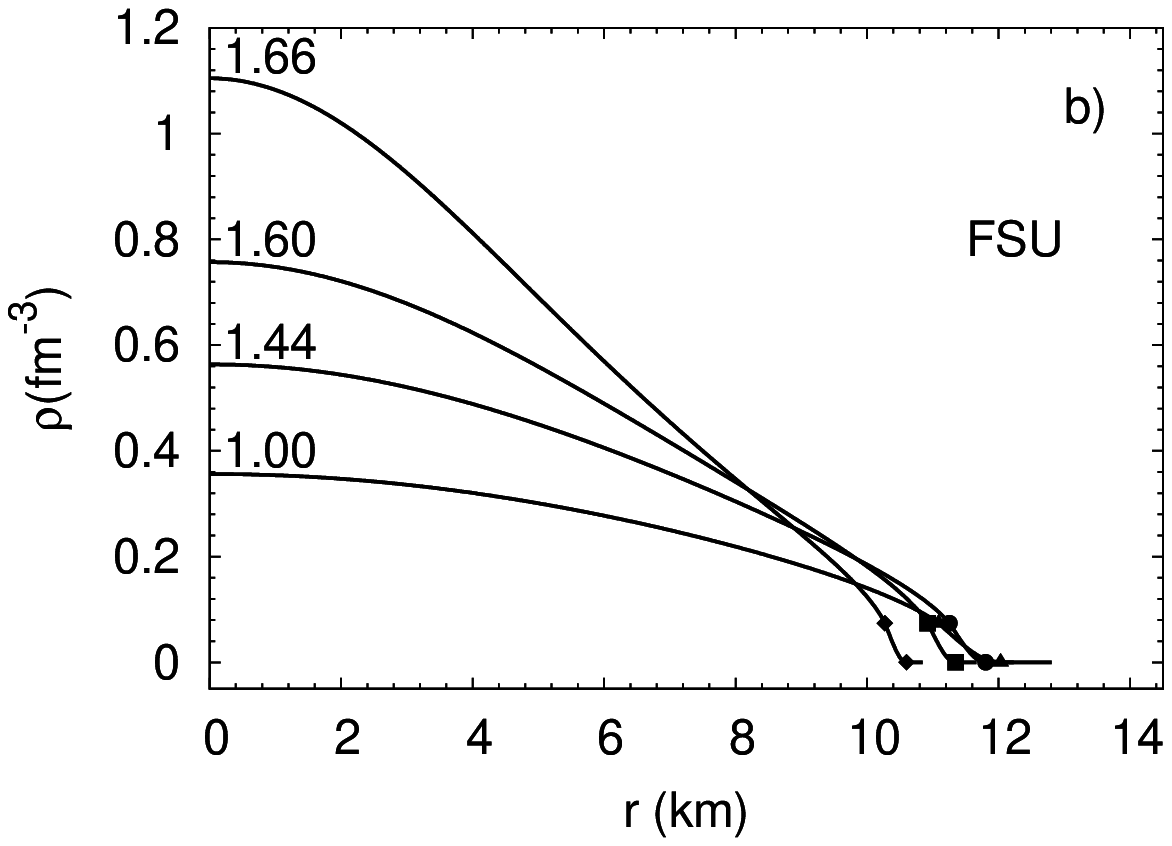}
  \includegraphics[angle=0,width=0.33\linewidth]{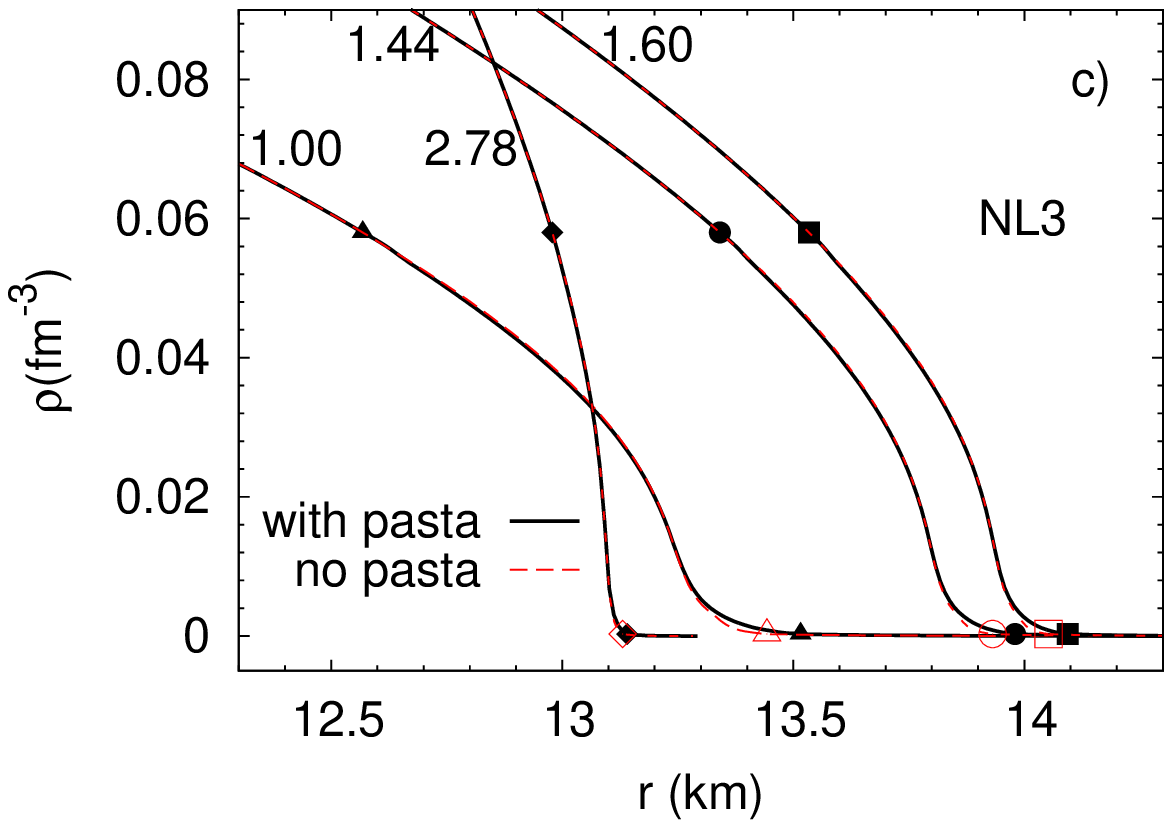}\\
  \includegraphics[angle=0,width=0.33\linewidth]{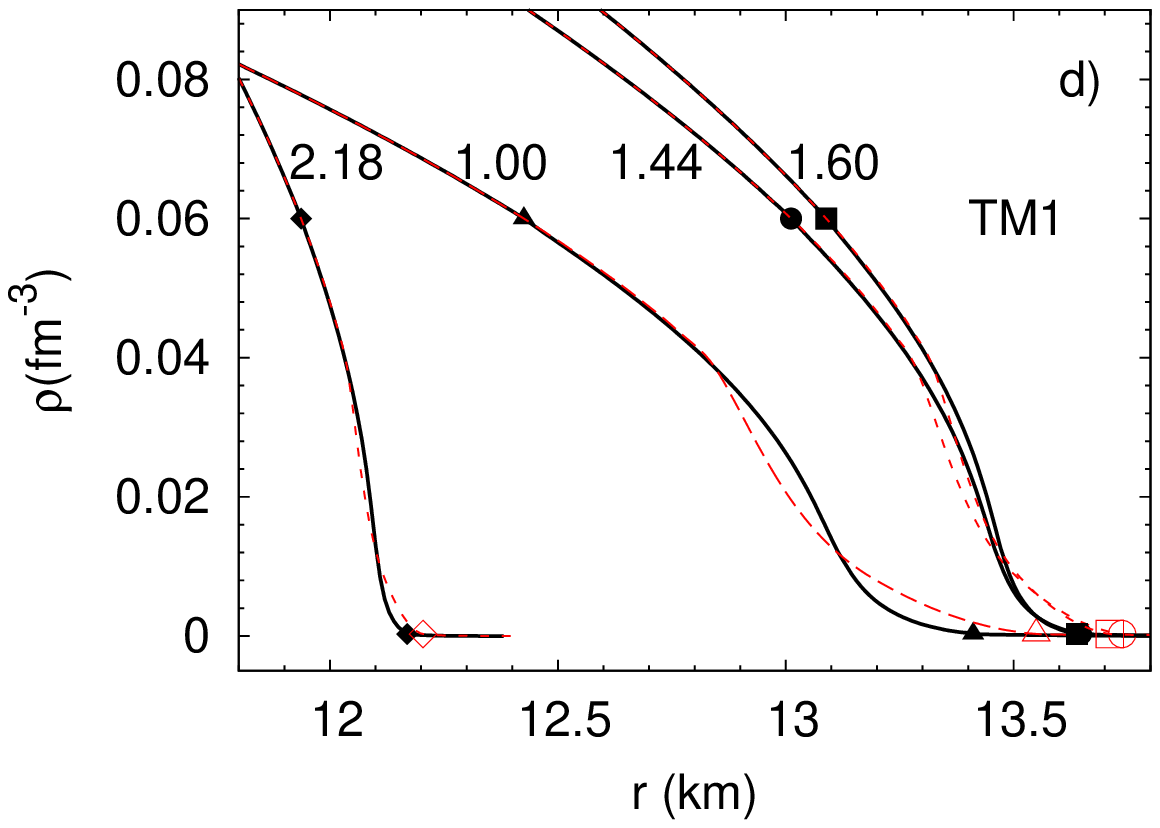}
  \includegraphics[angle=0,width=0.33\linewidth]{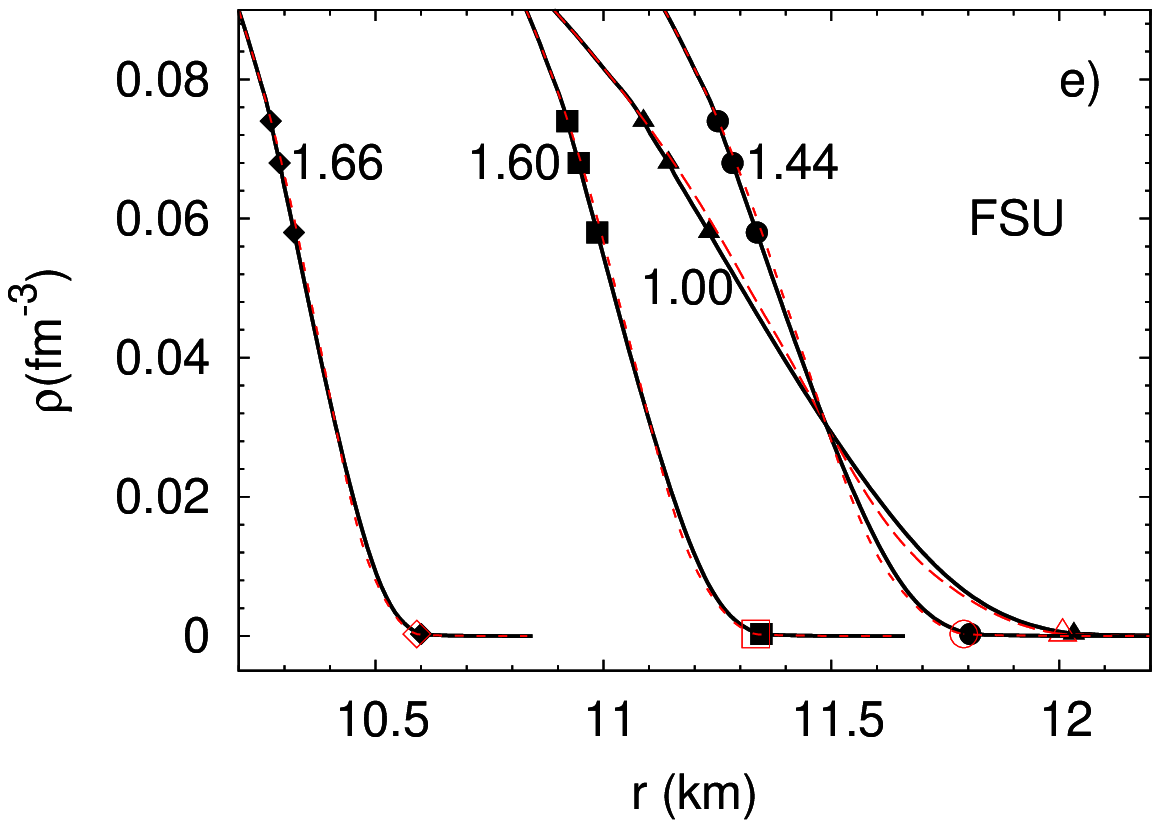}
  \includegraphics[angle=0,width=0.33\linewidth]{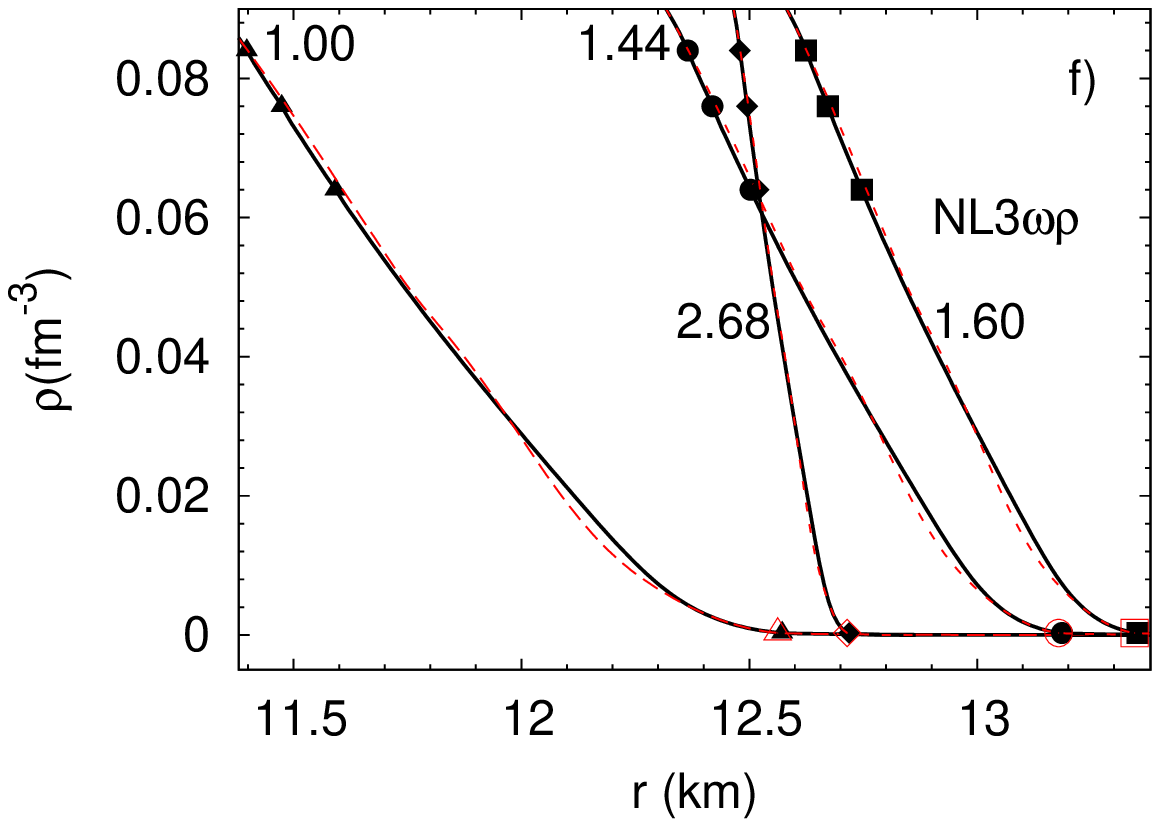}\\
  \includegraphics[angle=0,width=0.33\linewidth]{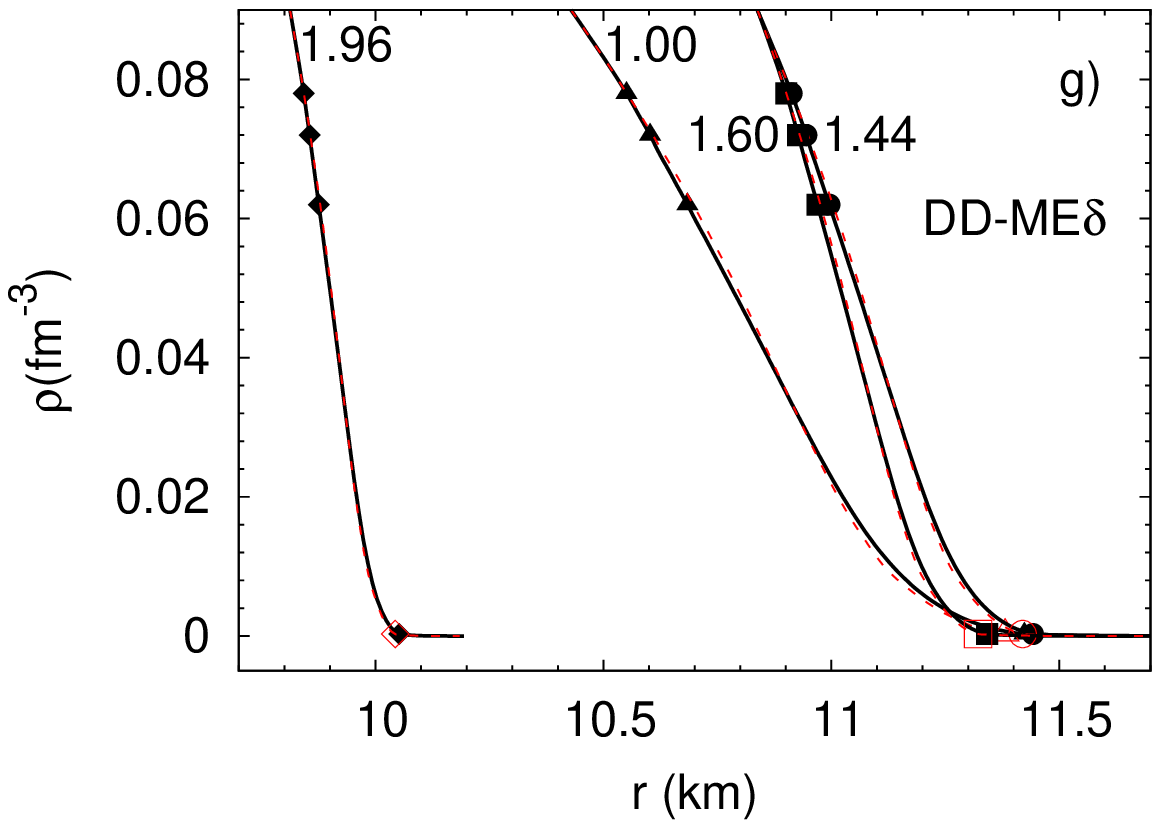}
  \includegraphics[angle=0,width=0.33\linewidth]{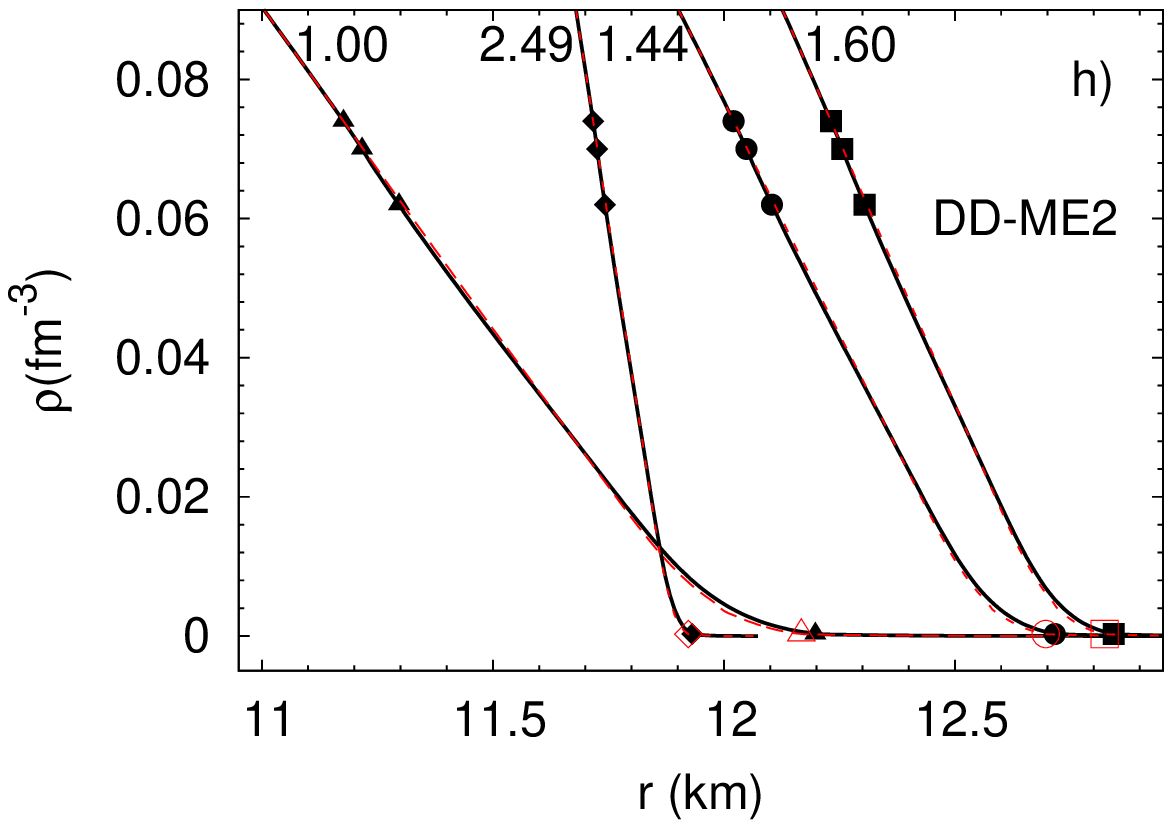}
  \includegraphics[angle=0,width=0.33\linewidth]{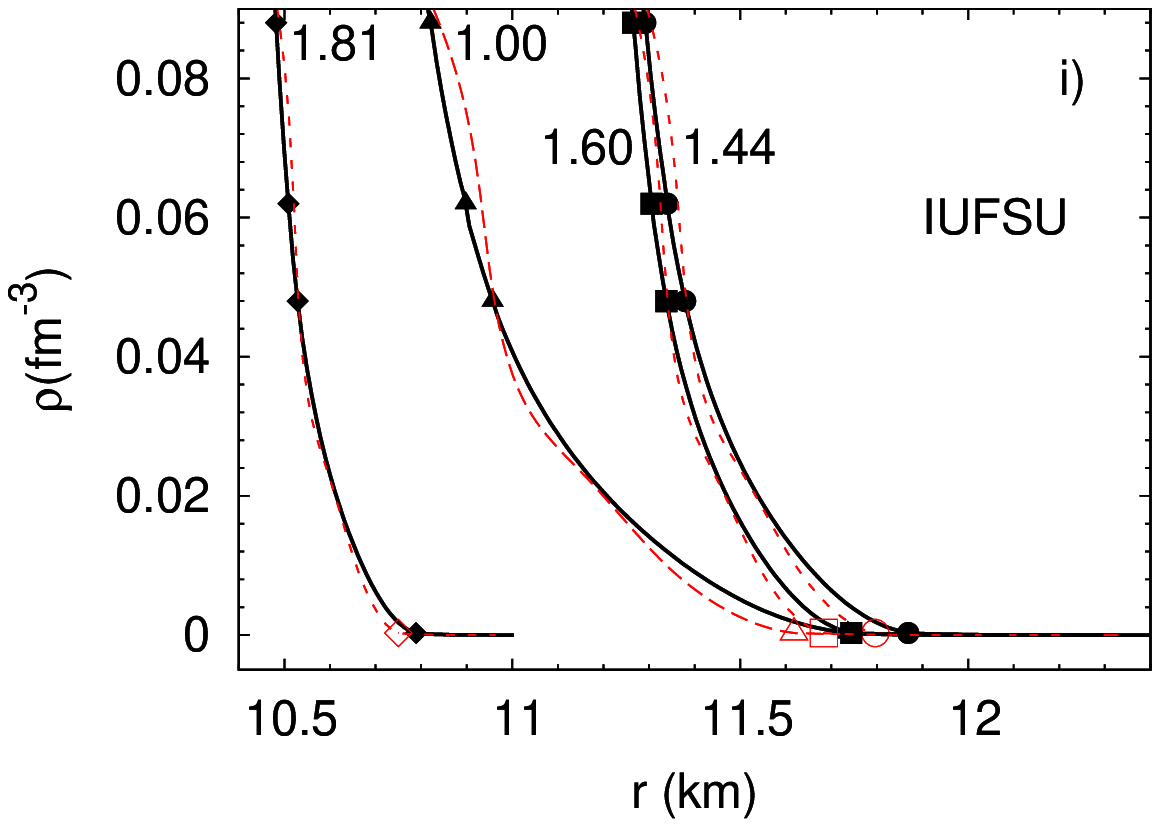}\\
  \end{tabular}
  \caption{(Color online) Profiles of neutron stars with a mass equal to 1.0 $M_\odot$,
    1.44 $M_\odot$, 1.6 $M_\odot$ and the maximum mass, for different nuclear models.
    The symbols stay at the the BPS-droplet, droplet-rod, rod-slab, slab-homogeneous transition
    for 1.0 $M_\odot$ (triangle),  1.44 $M_\odot$ (circle),  1.6 $M_\odot$ (square) and maximum mass
    (diamond). Figures a) and b) show the complete profile
    for NL3 and FSU, respectively. All other panels [c) to i)] focus on the crust profile.}
  \label{tov}
\end{figure*}

\subsection{Density profile of the crust}

In Fig. \ref{tov} we present the profile of stars with masses $M=1, 1.44, 1,60 M_\odot$
and $M_{max}$. The whole star profile is shown for the models NL3 and FSU in Figs.\ \ref{tov}a
and \ref{tov}b, respectively, since these two models predict the largest and the smallest maximum 
mass configuration.

\begin{figure}[tbh]
 \begin{tabular}{c}
  \includegraphics[width=0.8\linewidth]{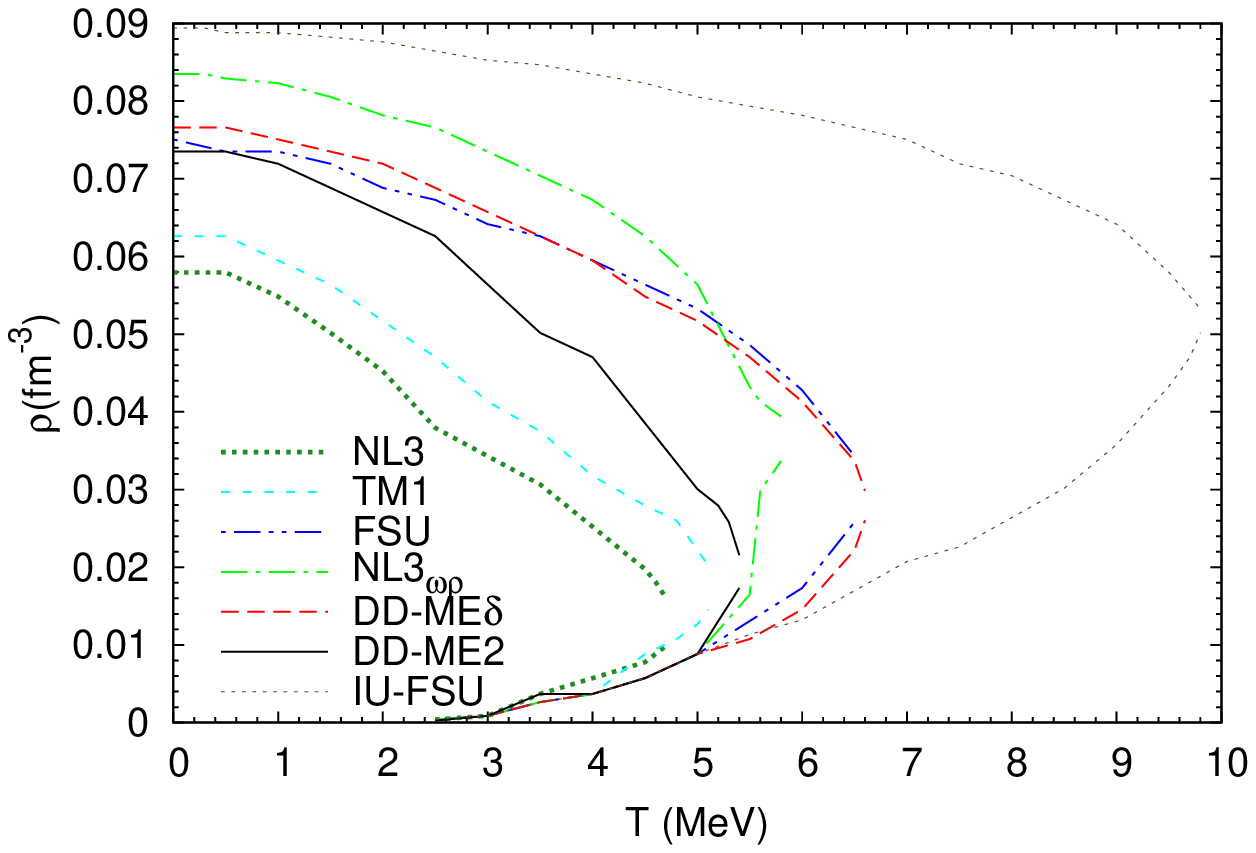}\\
  \includegraphics[width=0.8\linewidth]{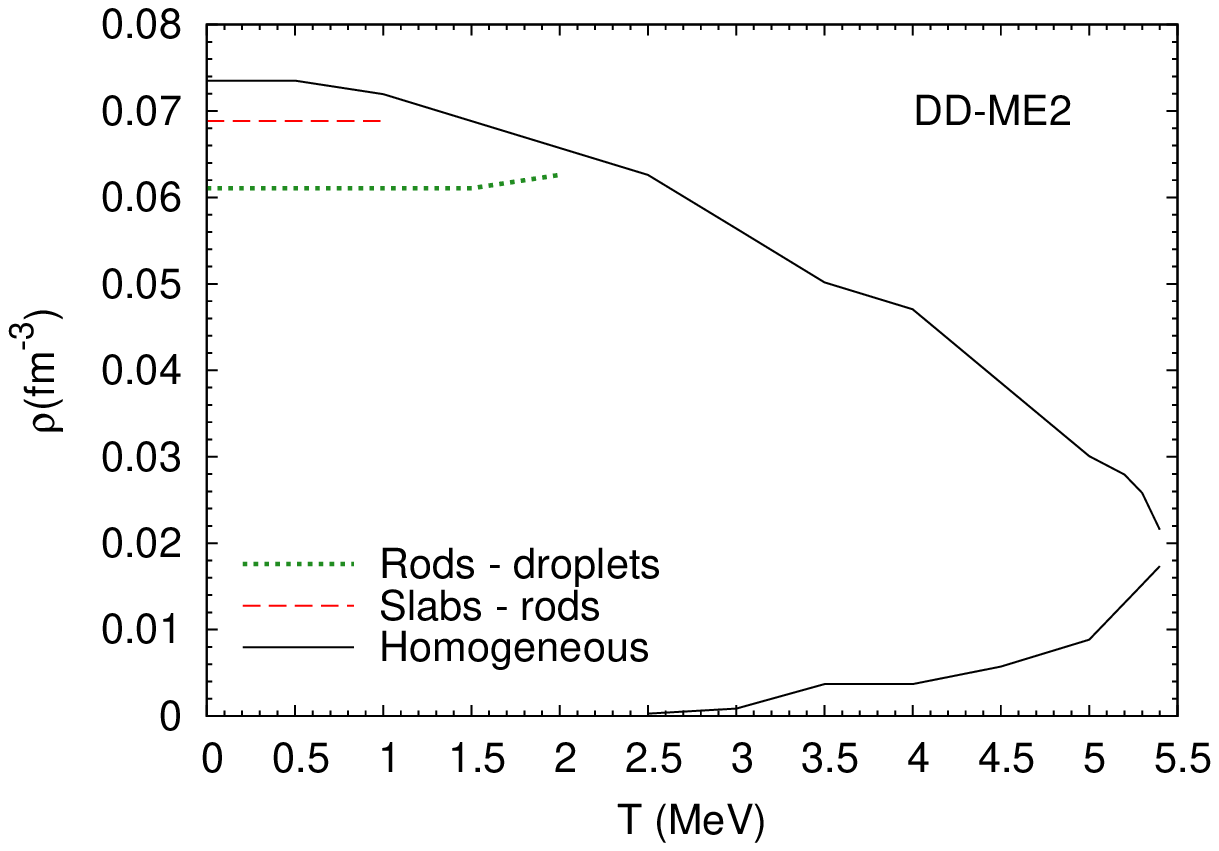}\\
  \includegraphics[width=0.8\linewidth]{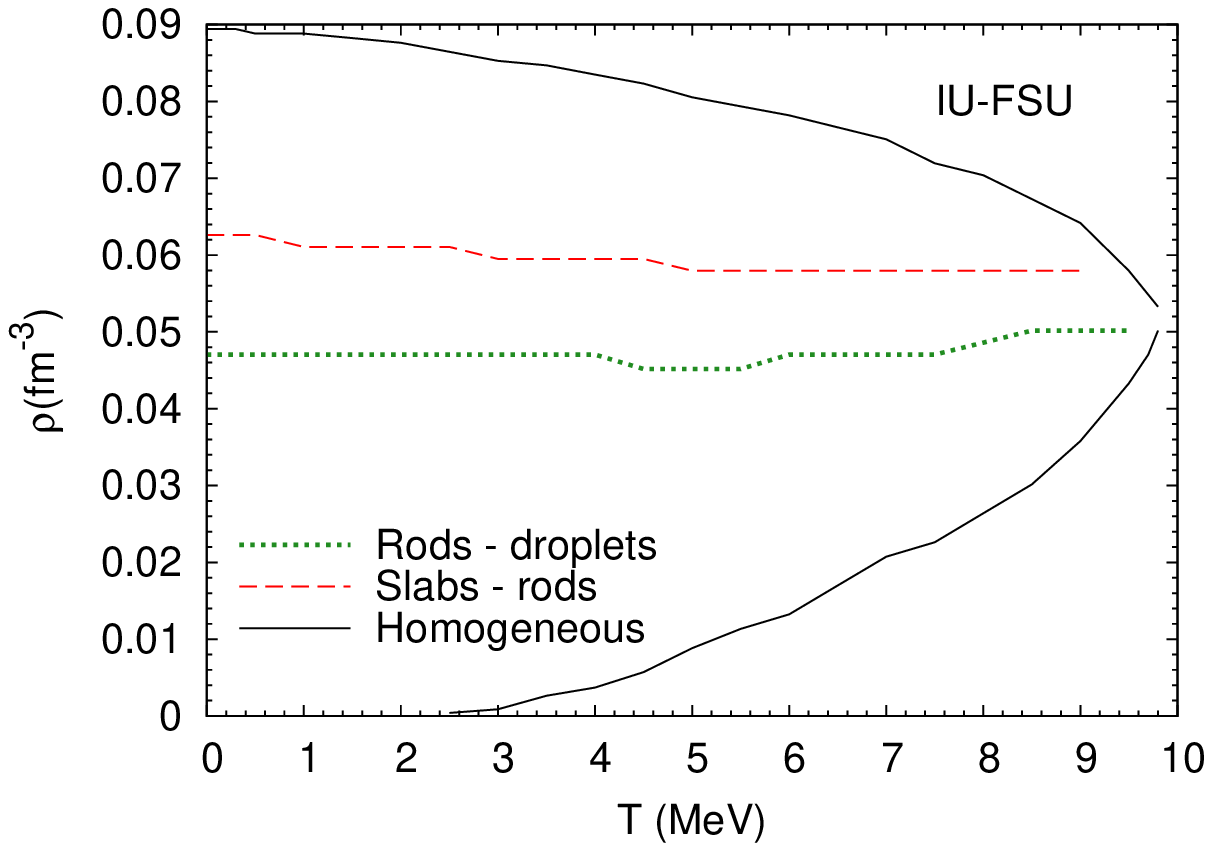}
 \end{tabular}
 \caption{(Color online) Top panel: density range of the crust as a function of temperature
   for all the models considered in the
   present study (top panel). Middle and bottom: size of the pasta phases
   versus $T$ for DDME2 and IU-FSU. In these two panels, the dotted (dashed) 
   line represents the transition
   droplet-rod (rod-slab).}
\label{tcrust}
\end{figure}

\begin{figure}[thb]
 \begin{tabular}{c}
  \includegraphics[width=0.8\linewidth]{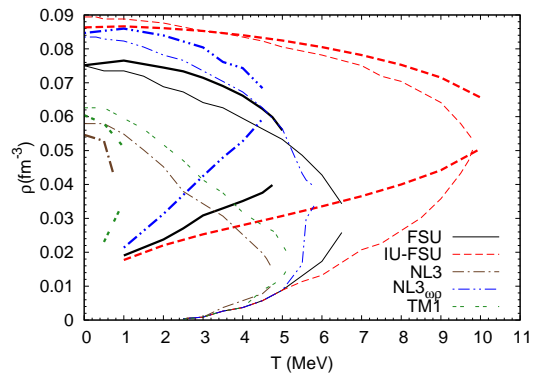}\\
    \end{tabular}
  \caption{(Color online) Density range of the crust as a function of temperature
    for some of the  relativistic mean field models considered in the
    present study (top panel), using a TF calculation (thin lines) and a DS calculation (thick lines).}
\label{tcrust1}
\end{figure}

All the other panels of Fig.\ \ref{tov} show, instead,  for the whole set of models under discussion,
only the last $\sim 2$ km of the star profile close to the surface.  The results determined with the same EoS 
presented in the previous section  ({\it i.e.,} including the TF calculation of the inner crust) are represented by a 
solid black line, and the transitions between the different phases of the inner crust are identified with black full symbols. 
For comparison, it is also shown the result obtained joining the BPS EoS directly to the homogeneous stellar matter EoS 
(red dashed lines). In this case, the transition from the BPS to the homogeneous matter is shown by a red empty symbol. 
The EoSs of the inner crust obtained within the TF framework and used to calculate the crust profiles are given in 
Tables \ref{eqState} and \ref{eqState2} of the Appendix.

Within the same model, a larger mass corresponds  to a steeper profile, as expected, due to the larger gravitational force.
In models with a large incompressibility, like  NL3, TM1, NL3$\omega\rho$ and DDME2, and taking only the 1, 1.44 and 1.6
$M_\odot$ stars, the star with the largest mass has the  inner crust at a larger distance from the center.  On the contrary, in the case of FSU, IU-FSU and DDME$\delta$ 
there is a larger concentration of mass at the center because the EoS is softer,  and the crust is          
pushed  more strongly {towards} the center of the star: this explains why
for IU-FSU and DDME$\delta$ the profiles of the 1.44 and 1.6
$M_\odot$ stars are almost
coincident, and for FSU the profiles of the 1.0 and 1.44 $M_\odot$ {stars}
cross, while the {crust} 
of the 1.6 $M_\odot$ {one} has the {smallest} distance to
the star center. Notice that for  NL3, TM1,
and DDME2 also the maximum mass star has the smallest radius.

 One interesting conclusion is that  taking into
 account  the  correct
 description of the inner crust in the total stellar EoS is more
 important for the softer EoS and with smaller slopes $L$. However, on the whole,
 using the BPS  EoS for the  outer crust and an EoS of homogeneous
 stellar matter for the inner crust and core gives good results { for
 the stellar profiles.}
\begin{table}[thb]
  \centering
  \begin{tabular}{l|ccccccc}
    \hline\hline
model & \multicolumn{2}{c}{ $\rho_{h-s}$ }    & $\rho_{s-r}$ &     $\rho_{r-d}$ &  $T_s$  &  $T_r$  &  $T_m$ \\
      & \multicolumn{2}{c}{ $(\mbox{fm}^{-3})$} & $(\mbox{fm}^{-3})$ & $(\mbox{fm}^{-3})$  &  (MeV)   &  (MeV)   &  (MeV) \\
\hline
	&					TF & DS	  &			&			&		&			&      \\	
    NL3                & $0.0579$ & 0.0546 &     -    &    -     &   -     &   -     & $4.7$  \\
    TM1                & $0.0626$ & 0.0604 &     -    &    -     &   -     &   -     & $5.1$  \\
    FSU                & $0.0751$ & 0.0751 & $0.0673$ & $0.0580$ & $2.5$   & $3.5$   & $6.5$  \\
    NL3$_{\omega\rho}$ & $0.0835$ & 0.0846 & $0.0751$ & $0.0642$ & $3.0$   & $3.5$   & $5.8$  \\
    DD-ME$\delta$      & $0.0766$ & - & $0.0720$ & $0.0626$ & $2.0$   & $3.0$   & $6.6$  \\
    DD-ME2             & $0.0735$ & - & $0.0688$ & $0.0611$ & $1.0$   & $2.0$   & $5.4$  \\
    IU-FSU             & $0.0894$ & 0.0863 & $0.0626$ & $0.0471$ & $9.0$   & $9.5$   & $9.8$  \\
    \hline\hline
  \end{tabular}
  \caption{Density transitions in the pasta phase at $T=0$, and melting
    temperature of the different pasta phases. For the transition to uniform matter we also show 
    the values obtained with a dynamical spinodal calculation (DS).}\label{TabTra}
\end{table}

\subsection{Finite temperature effects on the crust}

In this section we present the results of a finite temperature calculation of
the crust size for $\beta$-equilibrium matter. Our main objectives are: 
i) to determine the critical temperatures below which clusterization should be taken
into account; ii) to verify how temperature  affects the pasta phase, 
particularly the transition between the different geometries and the
size of the cells; and iii) to determine the melting temperature of clusters
with different geometry.

We perform the study using both a finite temperature TF
calculation of the pasta phase and the finite temperature dynamical
approach within the relativistic Vlasov formalism
\cite{cp2006,brito2006,pais}. In the latter case, the crust-core transition is
defined by the intersection between the $\beta$-equilibrium EoS and the
dynamical spinodal (DS); The cluster size is identified with half the wavelength of
the density fluctuation \cite{cp2006,brito2006,ducoin08}. The results are shown in 
Fig. \ref{tcrust} for the TF calculation; 
a comparison between the TF and the DS
calculation is done in Fig. \ref{tcrust1} and in Table \ref{TabTra}.

The critical temperature is model dependent and both the density dependence of
the symmetry energy and the incompressibility affect this quantity: a
smaller value of $L$ and a smaller $K_0$ favor clusterization at larger
temperatures. The results obtained within the TF
approach are compatible with the estimations calculated using the dynamical
calculation at the bottom of the inner crust. Close to the outer crust the
dynamical calculations estimations give much larger densities. This could be
expected, considering that in the TF calculation
the inner crust in this region is formed by small droplets inside cells with a
much larger radius. On the other hand, the dynamical calculation considers always that
the cluster size is half the WS cell, and therefore, has a much larger
surface energy contribution. 
Our results are compatible with the ones obtained in Ref.\ \cite{ducoin08}
(see Fig. 11 of this reference), where most of the results have been
obtained for Skyrme forces.

\begin{figure}[thb]
 \begin{tabular}{c}
  \includegraphics[width=0.7\linewidth]{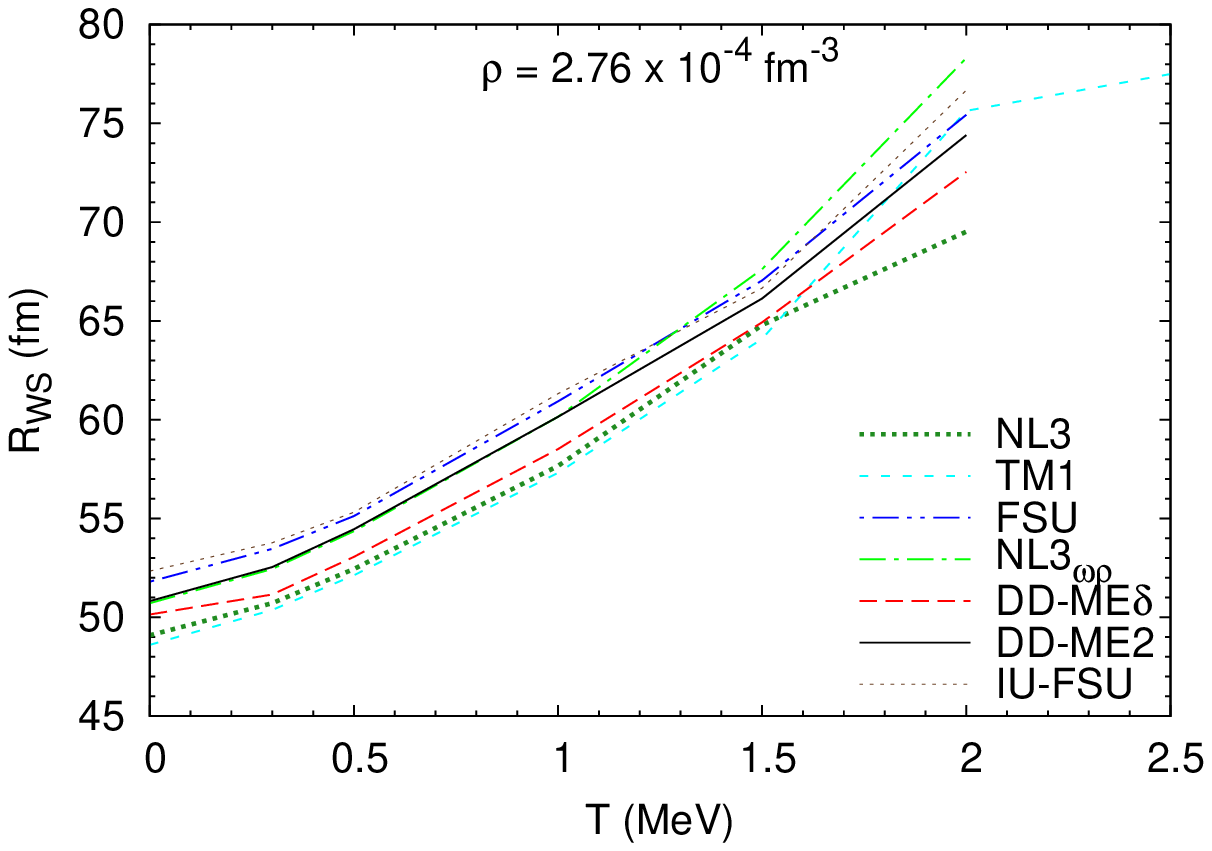}\\
  \includegraphics[width=0.7\linewidth]{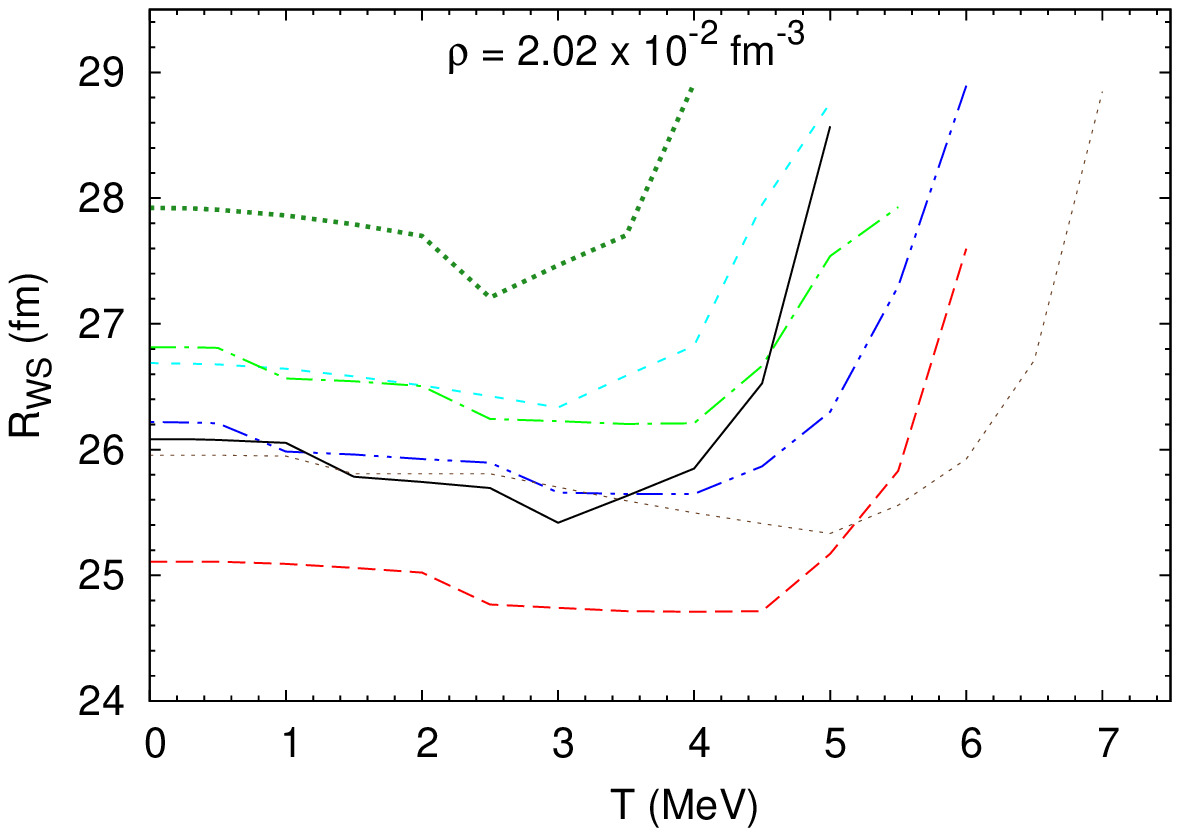}\\
  \includegraphics[width=0.7\linewidth]{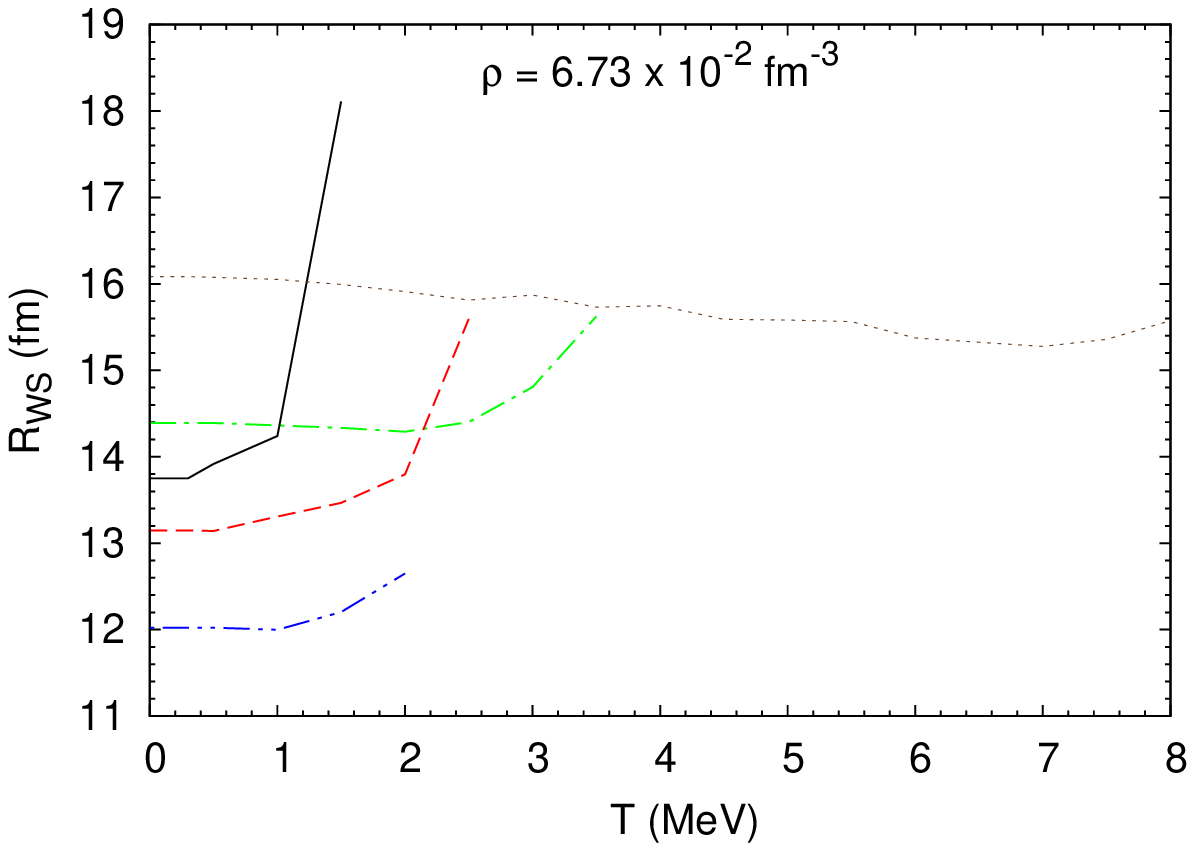}\\
  \end{tabular}
  \caption{(Color online)  Radius of the cells as a function of the
  temperature at the densities reported in top of the figures.} \label{radius}
\end{figure}

\begin{figure}[thb]
  \includegraphics[width=0.8\linewidth]{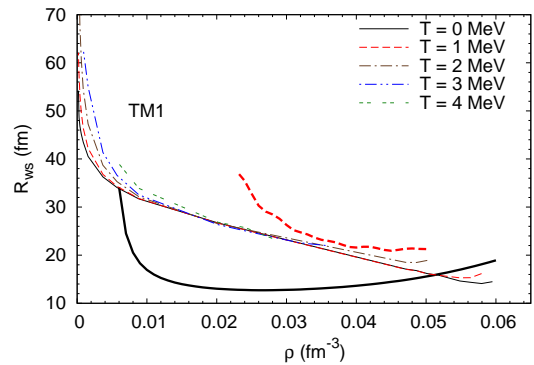}
  \caption{(Color online)  Radius of the cells as a function of the
  matter density for the TM1 model within the DS calculation (thick
  lines) and TF calculation (thin lines).} \label{radiusrho}
\end{figure}

From the middle and lower panels of Fig. \ref{tcrust}, 
we observe that the droplet-rod and the rod-slab transition 
densities do not depend on the temperature. However,
the melting temperature of the three geometries is different, with the
droplets surviving up to higher temperatures.

In the present calculation we suppose that the WS cells
exist until the clusters melt. This is an approximation that will
probably break down close to the melting point due to thermal
fluctuations, and therefore, the numbers obtained should be
interpreted as an upper limit.
 The problem of the effect of thermal fluctuations on the pasta
 structures has been studied in Refs. \cite{melt1,melt2} and it has been shown
 that thermally induced displacements of the rodlike and slablike
 nuclei can melt the lattice structure when these displacements are
 larger than the space available between the cluster and the boundary
 of the WS cell. Moreover, it was also shown that slablike
 nuclei would more easily be dissolved, while the rodlike were expected
 to survive at temperatures relevant for supernova cores.
In the present calculation, except for IU-FSU, all models predict the
melting of the slabs at a temperature $T< 3$ MeV, while the rods will melt at
temperatures one MeV higher, see Table \ref{TabTra}. Neglecting
thermal fluctuations we expect that clusters will survive in
$\beta$-equilibrium matter at temperatures below $5-6$ MeV,
approximately twice the melting temperature of non-droplet structures.

We next analyse the effect of the temperature on the size of the WS cells.
In Fig. \ref{radius} we plot the WS radius as a function of the temperature 
for all the models. We select three reference densities: $\rho=$2.76 $\times 10^{-4}$
and 2.02 $\times 10 ^{-2}$ fm$^{-3}$, at which the clusters are spherical for
all models, and $\rho=6.73 \times 10^{-2}$ fm$^{-3}$, where all are in
rod phase except IU-FSU, that is in a slab phase.
In Fig. \ref{radiusrho} we focus on the TM1 model, which only shows spherical 
clusters, and plot the radius of the WS cell versus density at different temperatures.

{
In general, for the densities shown,  the WS radius increases with the temperature, but for
$\rho=2.02 \times 10 ^{-2}$ fm$^{-3}$ the $R_{WS}$ suffers a small
decrease of the order of 0.5 fm below $T=3-4$ MeV. The surface energy
decreases  with temperature, and, as a result, we
could expect that the Wigner Seitz radius would decrease with
temperature. This, in fact, occurs for temperatures well below the critical transition 
temperature to homogeneous matter, corresponding to the behavior below
$T\lesssim 4$ MeV in the Fig. \ref{radius} middle panel for most of the models
considered, or to the IUFSU below 7 MeV and NL3$\omega\rho$ below 2 MeV in the bottom figure. 
 However, for temperatures
close to the critical temperature, the  Wigner Seitz radius increases most
probably due to the restrictions of the present calculation which does not
allow the freedom for the cluster to choose the shape that minimizes the free energy.
A similar behavior was obtained within a  different
formalism, namely considering plane-wave density fluctuations and relating the wave number of the unstable modes with the
Wigner Seitz cell size. In \cite{ducoin08} and \cite{pais}, both  in the
framework of the  non-relativistic Skyrme interactions
and the RMF nuclear models, the Wigner Seitz cell increases with the temperature. In
this case the size of the Wigner Seitz cell is identified with the
wave-length of the perturbation, and the cluster size with half-wave length,
and the system does not have the freedom to have the size of the cluster and
of the cell uncorrelated.
}

 We have performed a calculation of the  WS
radius within a DS calculation for TM1 and $T=0,$ and 1 MeV  (thick
lines in Fig. \ref{radiusrho}). TF and DS give  similar sizes for
$T=0$ MeV at the bottom of the inner crust,
where the dynamical calculation predicts non-homogeneous
matter. However, in general the DS predictions are quite different
from the TF results, in particular quite smaller at $T=0$ MeV and
larger at finite $T$.

\section{Conclusions}
\label{sec4}

In the present work the inner crust, including the non-spherical
pasta phases, was calculated within  a self-consistent Thomas--Fermi  
approach  \cite{pasta1,pasta2}  for $\beta$-equilibrium matter at zero and finite temperature. Several
relativistic nuclear models, both with  constant and density dependent
couplings, have been considered.

All models, except NL3 and TM1, both with a symmetry energy slope at saturation above 110 MeV, predict the existence of 
{\it lasagna}-like structures that may have an important contribution to the specific heat of the
crust \cite{luc2011}. 

The effect of the inner crust EoS on the neutron star profiles was also
analysed. It was verified that a smaller slope gives rise to a steeper crust {density profile} 
and a larger inner crust with respect to the total crust. It may also enhance
the slab phase size as observed in IU-FSU. 

It was observed that the star profile obtained using the TF inner crust
calculation or the homogeneous EoS above the outer-inner crust
transition did not differ much except for the models with a large
symmetry energy slope. 

The finite temperature calculation of the pasta phase in
$\beta$-equilibrium has shown that non-homogeneous matter is expected
for temperatures below 5-6 MeV, the only exception was  obtained with
the  parametrization IU-FSU which only melts at 9.5
MeV. Non-spherical structures, rodlike and slablike, melt above,
respectively, 2-3.5 MeV and 1-3 MeV. It was also verified that the
onset density of the rodlike and  the slablike structures is independent of
the temperature.

{Finally, it was shown that a DS calculation is able to give a good
prediction of the crust-core transition, even at finite
temperature. However, considering the size of the WS cells, this
formalism fails except for $T=0$ MeV close to the crust-core transition.}

\section*{Acknowledgments}
We are grateful to {Silvia Chiacchiera} for the careful reading of
the manuscript and for the editing of the figures.
This work has been partially supported by QREN/FEDER, the
Programme COMPETE, and FCT (Portugal) under the
projects PTDC/FIS/113292/2009 and CERN/FP/123608/2011, by the Capes/FCT n. 232/09 bilateral
collaboration.


\section*{Appendix: Equation of state of the inner crust}
\label{app1}

\begin{table*}[tbh]
 \caption{Equation of state of the inner crust for the seven model considered. The energy density is given in unit of $10$ fm$^{-4}$, while the pressure is in unit of $10^3$ fm$^{-4}$.}\label{eqState}
  \centering
  \footnotesize{
  \begin{tabular}{c|rr|rr|rr|rr|rr|rr|rr}
\hline\hline
             &   \multicolumn{2}{c|}{NL3}  &\multicolumn{2}{c|}{TM1}&\multicolumn{2}{c|}{FSU}&\multicolumn{2}{c|}{NL3$_{\omega\rho}$}&\multicolumn{2}{c|}{DD-ME$\delta$}& \multicolumn{2}{c|}{DD-ME2}  &\multicolumn{2}{c}{IU-FSU}\\
$\rho$ (fm$^{-3}$)  &  \multicolumn{1}{c}{$\mathcal E$} & \multicolumn{1}{c|}{P}  &  \multicolumn{1}{c}{$\mathcal E$} & \multicolumn{1}{c|}{P}  &  \multicolumn{1}{c}{$\mathcal E$} & \multicolumn{1}{c|}{P}  &  \multicolumn{1}{c}{$\mathcal E$} & \multicolumn{1}{c|}{P}
&  \multicolumn{1}{c}{$\mathcal E$} & \multicolumn{1}{c|}{P}  &  \multicolumn{1}{c}{$\mathcal E$} & \multicolumn{1}{c|}{P}  &  \multicolumn{1}{c}{$\mathcal E$} & \multicolumn{1}{c}{P} \\
\hline
 $0.0860$    &      -      &      -      &      -      &      -        &      -      &      -        &      -      &      -        &      -      &      -        &      -      &      -        & $ 4.13948$  & $ 1.60166$ \\
 $0.0840$    &      -      &      -      &      -      &      -        &      -      &      -        &      -      &      -        &      -      &      -        &      -      &      -        & $ 4.04284$  & $ 1.57348$ \\
 $0.0820$    &      -      &      -      &      -      &      -        &      -      &      -        & $ 3.94154$  & $ 2.52403$    &      -      &      -        &      -      &      -        & $ 3.94621$  & $ 1.54348$ \\
 $0.0800$    &      -      &      -      &      -      &      -        &      -      &      -        & $ 3.84480$  & $ 2.41543$    &      -      &      -        &      -      &      -        & $ 3.84959$  & $ 1.50902$ \\
 $0.0780$    &      -      &      -      &      -      &      -        &      -      &      -        & $ 3.74809$  & $ 2.30744$    &      -      &      -        &      -      &      -        & $ 3.75297$  & $ 1.47400$ \\
 $0.0760$    &      -      &      -      &      -      &      -        &      -      &      -        & $ 3.65141$  & $ 2.20122$    & $ 3.65126$  & $ 2.04088$    &      -      &      -        & $ 3.65637$  & $ 1.43842$ \\
 $0.0740$    &      -      &      -      &      -      &      -        &      -      &      -        & $ 3.55475$  & $ 2.11172$    & $ 3.55465$  & $ 1.93136$    &      -      &      -        & $ 3.55977$  & $ 1.40239$ \\
 $0.0720$    &      -      &      -      &      -      &      -        & $ 3.45832$  & $ 1.86234$    & $ 3.45812$  & $ 2.01229$    & $ 3.45807$  & $ 1.82398$    & $ 3.45856$  & $ 2.06135$    & $ 3.46319$  & $ 1.36591$ \\
 $0.0700$    &      -      &      -      &      -      &      -        & $ 3.36175$  & $ 1.76038$    & $ 3.36151$  & $ 1.91484$    & $ 3.36152$  & $ 1.73271$    & $ 3.36193$  & $ 1.95675$    & $ 3.36661$  & $ 1.32916$ \\
 $0.0680$    &      -      &      -      &      -      &      -        & $ 3.26521$  & $ 1.66110$    & $ 3.26494$  & $ 1.81906$    & $ 3.26500$  & $ 1.63419$    & $ 3.26533$  & $ 1.86477$    & $ 3.27005$  & $ 1.29192$ \\
 $0.0660$    &      -      &      -      &      -      &      -        & $ 3.16870$  & $ 1.57890$    & $ 3.16839$  & $ 1.72541$    & $ 3.16850$  & $ 1.53902$    & $ 3.16875$  & $ 1.76524$    & $ 3.17350$  & $ 1.25411$ \\
 $0.0640$    &      -      &      -      &      -      &      -        & $ 3.07221$  & $ 1.48951$    & $ 3.07187$  & $ 1.63348$    & $ 3.07203$  & $ 1.44694$    & $ 3.07221$  & $ 1.66744$    & $ 3.07696$  & $ 1.21646$ \\
 $0.0620$    &      -      &      -      &      -      &      -        & $ 2.97575$  & $ 1.40336$    & $ 2.97537$  & $ 1.54804$    & $ 2.97559$  & $ 1.35825$    & $ 2.97570$  & $ 1.57160$    & $ 2.98043$  & $ 1.17734$ \\
 $0.0600$    &      -      &      -      &      -      &      -        & $ 2.87932$  & $ 1.31999$    & $ 2.87891$  & $ 1.46017$    & $ 2.87918$  & $ 1.27656$    & $ 2.87922$  & $ 1.48170$    & $ 2.88391$  & $ 1.15722$ \\
 $0.0580$    &      -      &      -      & $ 2.77736$  & $ 1.48251$    & $ 2.78292$  & $ 1.24043$    & $ 2.78247$  & $ 1.37452$    & $ 2.78280$  & $ 1.19477$    & $ 2.78276$  & $ 1.39043$    & $ 2.78740$  & $ 1.11951$ \\
 $0.0560$    & $ 2.67959$  & $ 1.36758$  & $ 2.68110$  & $ 1.33925$    & $ 2.68654$  & $ 1.16654$    & $ 2.68607$  & $ 1.29055$    & $ 2.68644$  & $ 1.11632$    & $ 2.68634$  & $ 1.30149$    & $ 2.69090$  & $ 1.08100$ \\
 $0.0540$    & $ 2.58343$  & $ 1.24190$  & $ 2.58489$  & $ 1.20460$    & $ 2.59019$  & $ 1.09281$    & $ 2.58969$  & $ 1.20876$    & $ 2.59011$  & $ 1.04117$    & $ 2.58995$  & $ 1.21468$    & $ 2.59442$  & $ 1.04177$ \\
 $0.0520$    & $ 2.48731$  & $ 1.10988$  & $ 2.48873$  & $ 1.07826$    & $ 2.49387$  & $ 1.02186$    & $ 2.49334$  & $ 1.12899$    & $ 2.49381$  & $ 0.96925$    & $ 2.49359$  & $ 1.13056$    & $ 2.49795$  & $ 1.00189$ \\
 $0.0500$    & $ 2.39124$  & $ 0.98623$  & $ 2.39262$  & $ 0.96135$    & $ 2.39757$  & $ 0.95359$    & $ 2.39703$  & $ 1.05140$    & $ 2.39753$  & $ 0.90079$    & $ 2.39727$  & $ 1.04932$    & $ 2.40150$  & $ 0.96104$ \\
 $0.0480$    & $ 2.29522$  & $ 0.87140$  & $ 2.29656$  & $ 0.85336$    & $ 2.30130$  & $ 0.88792$    & $ 2.30074$  & $ 0.97589$    & $ 2.30129$  & $ 0.83526$    & $ 2.30097$  & $ 0.97047$    & $ 2.30506$  & $ 0.91959$ \\
 $0.0460$    & $ 2.19925$  & $ 0.76548$  & $ 2.20053$  & $ 0.75398$    & $ 2.20505$  & $ 0.82493$    & $ 2.20448$  & $ 0.90271$    & $ 2.20506$  & $ 0.77288$    & $ 2.20471$  & $ 0.89456$    & $ 2.20864$  & $ 0.88260$ \\
 $0.0440$    & $ 2.10332$  & $ 0.66838$  & $ 2.10455$  & $ 0.66326$    & $ 2.10884$  & $ 0.76421$    & $ 2.10826$  & $ 0.83187$    & $ 2.10887$  & $ 0.71384$    & $ 2.10848$  & $ 0.82153$    & $ 2.11224$  & $ 0.83977$ \\
 $0.0420$    & $ 2.00743$  & $ 0.57990$  & $ 2.00861$  & $ 0.58036$    & $ 2.01265$  & $ 0.70583$    & $ 2.01207$  & $ 0.76345$    & $ 2.01270$  & $ 0.65749$    & $ 2.01228$  & $ 0.75175$    & $ 2.01586$  & $ 0.79629$ \\
$0.0400$    & $ 1.91158$  & $ 0.49988$  & $ 1.91270$  & $ 0.50581$    & $ 1.91648$  & $ 0.65009$    & $ 1.91591$  & $ 0.69757$    & $ 1.91656$  & $ 0.60433$    & $ 1.91612$  & $ 0.68485$    & $ 1.91950$  & $ 0.75215$ \\
 $0.0380$    & $ 1.81577$  & $ 0.42802$  & $ 1.81683$  & $ 0.43892$    & $ 1.82035$  & $ 0.59632$    & $ 1.81978$  & $ 0.63438$    & $ 1.82044$  & $ 0.55375$    & $ 1.81999$  & $ 0.62125$    & $ 1.82316$  & $ 0.70710$ \\
 $0.0360$    & $ 1.71999$  & $ 0.36366$  & $ 1.72099$  & $ 0.37922$    & $ 1.72424$  & $ 0.54503$    & $ 1.72368$  & $ 0.57417$    & $ 1.72435$  & $ 0.50611$    & $ 1.72389$  & $ 0.56110$    & $ 1.72684$  & $ 0.66164$ \\
 $0.0340$    & $ 1.62425$  & $ 0.30710$  & $ 1.62519$  & $ 0.32641$    & $ 1.62816$  & $ 0.49573$    & $ 1.62762$  & $ 0.51660$    & $ 1.62828$  & $ 0.46111$    & $ 1.62782$  & $ 0.50414$    & $ 1.63055$  & $ 0.61573$ \\
 $0.0320$    & $ 1.52854$  & $ 0.25764$  & $ 1.52941$  & $ 0.27974$    & $ 1.53211$  & $ 0.44880$    & $ 1.53159$  & $ 0.46202$    & $ 1.53224$  & $ 0.41839$    & $ 1.53179$  & $ 0.45077$    & $ 1.53429$  & $ 0.56936$ \\
 $0.0300$    & $ 1.43286$  & $ 0.21487$  & $ 1.43366$  & $ 0.23935$    & $ 1.43609$  & $ 0.40395$    & $ 1.43560$  & $ 0.41054$    & $ 1.43623$  & $ 0.37826$    & $ 1.43578$  & $ 0.40065$    & $ 1.43805$  & $ 0.52279$ \\
 $0.0280$    & $ 1.33721$  & $ 0.17828$  & $ 1.33793$  & $ 0.20443$    & $ 1.34009$  & $ 0.36143$    & $ 1.33963$  & $ 0.36219$    & $ 1.34024$  & $ 0.34020$    & $ 1.33981$  & $ 0.35413$    & $ 1.34185$  & $ 0.47616$ \\
 $0.0260$    & $ 1.24158$  & $ 0.14742$  & $ 1.24223$  & $ 0.17453$    & $ 1.24413$  & $ 0.32094$    & $ 1.24370$  & $ 0.31724$    & $ 1.24428$  & $ 0.30452$    & $ 1.24388$  & $ 0.31080$    & $ 1.24568$  & $ 0.42990$ \\
 $0.0240$    & $ 1.14597$  & $ 0.12168$  & $ 1.14655$  & $ 0.14904$    & $ 1.14819$  & $ 0.28293$    & $ 1.14781$  & $ 0.27538$    & $ 1.14835$  & $ 0.27072$    & $ 1.14797$  & $ 0.27102$    & $ 1.14955$  & $ 0.38403$ \\
 $0.0220$    & $ 1.05038$  & $ 0.10070$  & $ 1.05089$  & $ 0.12750$    & $ 1.05229$  & $ 0.24695$    & $ 1.05194$  & $ 0.23687$    & $ 1.05244$  & $ 0.23894$    & $ 1.05210$  & $ 0.23433$    & $ 1.05345$  & $ 0.33888$ \\
 $0.0200$    & $ 0.95481$  & $ 0.08382$  & $ 0.95525$  & $ 0.10921$    & $ 0.95642$  & $ 0.21340$    & $ 0.95611$  & $ 0.20175$    & $ 0.95656$  & $ 0.20879$    & $ 0.95625$  & $ 0.20094$    & $ 0.95740$  & $ 0.29484$ \\
 $0.0180$    & $ 0.85925$  & $ 0.07034$  & $ 0.85962$  & $ 0.09385$    & $ 0.86058$  & $ 0.18198$    & $ 0.86032$  & $ 0.16967$    & $ 0.86071$  & $ 0.18046$    & $ 0.86044$  & $ 0.17033$    & $ 0.86138$  & $ 0.25222$ \\
 $0.0160$    & $ 0.76371$  & $ 0.05970$  & $ 0.76401$  & $ 0.08068$    & $ 0.76478$  & $ 0.15294$    & $ 0.76455$  & $ 0.14068$    & $ 0.76489$  & $ 0.15360$    & $ 0.76467$  & $ 0.14261$    & $ 0.76542$  & $ 0.21158$ \\
 $0.0140$    & $ 0.66817$  & $ 0.05118$  & $ 0.66842$  & $ 0.06917$    & $ 0.66901$  & $ 0.12598$    & $ 0.66883$  & $ 0.11489$    & $ 0.66910$  & $ 0.12837$    & $ 0.66892$  & $ 0.11747$    & $ 0.66950$  & $ 0.17301$ \\
 $0.0120$    & $ 0.57265$  & $ 0.04399$  & $ 0.57284$  & $ 0.05873$    & $ 0.57327$  & $ 0.10120$    & $ 0.57313$  & $ 0.09162$    & $ 0.57335$  & $ 0.10445$    & $ 0.57321$  & $ 0.09461$    & $ 0.57364$  & $ 0.13703$ \\
     \hline\hline
  \end{tabular}}
 \end{table*}

\begin{table*}[tbh]
  \caption{(Continuation).}
\label{eqState2}
  \centering
  \footnotesize{
  \begin{tabular}{c|rr|rr|rr|rr|rr|rr|rr}
\hline\hline
             &   \multicolumn{2}{c|}{NL3}  &\multicolumn{2}{c|}{TM1}&\multicolumn{2}{c|}{FSU}&\multicolumn{2}{c|}{NL3$_{\omega\rho}$}&\multicolumn{2}{c|}{DD-ME$\delta$}& \multicolumn{2}{c|}{DD-ME2}  &\multicolumn{2}{c}{IU-FSU}\\
$\rho$ (fm$^{-3}$)  &  \multicolumn{1}{c}{$\mathcal E$} & \multicolumn{1}{c|}{P}  &  \multicolumn{1}{c}{$\mathcal E$} & \multicolumn{1}{c|}{P}  &  \multicolumn{1}{c}{$\mathcal E$} & \multicolumn{1}{c|}{P}  &  \multicolumn{1}{c}{$\mathcal E$} & \multicolumn{1}{c|}{P}
&  \multicolumn{1}{c}{$\mathcal E$} & \multicolumn{1}{c|}{P}  &  \multicolumn{1}{c}{$\mathcal E$} & \multicolumn{1}{c|}{P}  &  \multicolumn{1}{c}{$\mathcal E$} & \multicolumn{1}{c}{P} \\
\hline

 $0.0100$    & $ 0.47714$  & $ 0.03760$  & $ 0.47697$  & $ 0.04799$    & $ 0.47758$  & $ 0.07870$    & $ 0.47748$  & $ 0.07095$    & $ 0.47764$  & $ 0.08210$    & $ 0.47754$  & $ 0.07389$    & $ 0.47783$  & $ 0.10429$ \\
 $0.0095$    & $ 0.45327$  & $ 0.03608$  & $ 0.45310$  & $ 0.04561$    & $ 0.45366$  & $ 0.07338$    & $ 0.45357$  & $ 0.06618$    & $ 0.45372$  & $ 0.07673$    & $ 0.45362$  & $ 0.06897$    & $ 0.45389$  & $ 0.09659$ \\
 $0.0090$    & $ 0.42939$  & $ 0.03451$  & $ 0.42923$  & $ 0.04318$    & $ 0.42975$  & $ 0.06816$    & $ 0.42966$  & $ 0.06157$    & $ 0.42980$  & $ 0.07145$    & $ 0.42971$  & $ 0.06421$    & $ 0.42995$  & $ 0.08909$ \\
 $0.0085$    & $ 0.40552$  & $ 0.03289$  & $ 0.40536$  & $ 0.04074$    & $ 0.40584$  & $ 0.06309$    & $ 0.40576$  & $ 0.05706$    & $ 0.40588$  & $ 0.06624$    & $ 0.40581$  & $ 0.05955$    & $ 0.40602$  & $ 0.08184$ \\
 $0.0080$    & $ 0.38165$  & $ 0.03132$  & $ 0.38149$  & $ 0.03836$    & $ 0.38193$  & $ 0.05818$    & $ 0.38186$  & $ 0.05270$    & $ 0.38197$  & $ 0.06122$    & $ 0.38190$  & $ 0.05504$    & $ 0.38209$  & $ 0.07485$ \\
 $0.0075$    & $ 0.35777$  & $ 0.02970$  & $ 0.35763$  & $ 0.03588$    & $ 0.35802$  & $ 0.05336$    & $ 0.35796$  & $ 0.04840$    & $ 0.35806$  & $ 0.05620$    & $ 0.35800$  & $ 0.05058$    & $ 0.35816$  & $ 0.06806$ \\
 $0.0070$    & $ 0.33390$  & $ 0.02797$  & $ 0.33376$  & $ 0.03340$    & $ 0.33412$  & $ 0.04870$    & $ 0.33407$  & $ 0.04429$    & $ 0.33415$  & $ 0.05139$    & $ 0.33410$  & $ 0.04632$    & $ 0.33424$  & $ 0.06152$ \\
 $0.0065$    & $ 0.31003$  & $ 0.02630$  & $ 0.30990$  & $ 0.03096$    & $ 0.31022$  & $ 0.04414$    & $ 0.31017$  & $ 0.04024$    & $ 0.31025$  & $ 0.04657$    & $ 0.31021$  & $ 0.04211$    & $ 0.31033$  & $ 0.05529$ \\
 $0.0060$    & $ 0.28617$  & $ 0.02453$  & $ 0.28604$  & $ 0.02853$    & $ 0.28633$  & $ 0.03978$    & $ 0.28629$  & $ 0.03639$    & $ 0.28635$  & $ 0.04196$    & $ 0.28631$  & $ 0.03801$    & $ 0.28642$  & $ 0.04916$ \\
 $0.0055$    & $ 0.26230$  & $ 0.02270$  & $ 0.26218$  & $ 0.02605$    & $ 0.26243$  & $ 0.03552$    & $ 0.26240$  & $ 0.03264$    & $ 0.26246$  & $ 0.03750$    & $ 0.26242$  & $ 0.03406$    & $ 0.26251$  & $ 0.04338$ \\
 $0.0050$    & $ 0.23843$  & $ 0.02088$  & $ 0.23832$  & $ 0.02351$    & $ 0.23855$  & $ 0.03137$    & $ 0.23852$  & $ 0.02899$    & $ 0.23856$  & $ 0.03309$    & $ 0.23854$  & $ 0.03015$    & $ 0.23861$  & $ 0.03791$ \\
 $0.0045$    & $ 0.21457$  & $ 0.01895$  & $ 0.21447$  & $ 0.02108$    & $ 0.21466$  & $ 0.02742$    & $ 0.21464$  & $ 0.02544$    & $ 0.21468$  & $ 0.02889$    & $ 0.21466$  & $ 0.02645$    & $ 0.21471$  & $ 0.03269$ \\
 $0.0040$    & $ 0.19071$  & $ 0.01703$  & $ 0.19061$  & $ 0.01855$    & $ 0.19078$  & $ 0.02362$    & $ 0.19076$  & $ 0.02204$    & $ 0.19079$  & $ 0.02488$    & $ 0.19078$  & $ 0.02286$    & $ 0.19082$  & $ 0.02777$ \\
 $0.0035$    & $ 0.16685$  & $ 0.01500$  & $ 0.16677$  & $ 0.01606$    & $ 0.16691$  & $ 0.02002$    & $ 0.16689$  & $ 0.01880$    & $ 0.16692$  & $ 0.02098$    & $ 0.16690$  & $ 0.01941$    & $ 0.16694$  & $ 0.02316$ \\
 $0.0030$    & $ 0.14299$  & $ 0.01297$  & $ 0.14292$  & $ 0.01363$    & $ 0.14304$  & $ 0.01657$    & $ 0.14303$  & $ 0.01571$    & $ 0.14304$  & $ 0.01733$    & $ 0.14304$  & $ 0.01612$    & $ 0.14306$  & $ 0.01880$ \\
 $0.0025$    & $ 0.11914$  & $ 0.01095$  & $ 0.11908$  & $ 0.01125$    & $ 0.11917$  & $ 0.01333$    & $ 0.11916$  & $ 0.01272$    & $ 0.11918$  & $ 0.01389$    & $ 0.11917$  & $ 0.01297$    & $ 0.11919$  & $ 0.01485$ \\
 $0.0020$    & $ 0.09529$  & $ 0.00892$  & $ 0.09524$  & $ 0.00892$    & $ 0.09532$  & $ 0.01034$    & $ 0.09531$  & $ 0.00998$    & $ 0.09532$  & $ 0.01069$    & $ 0.09531$  & $ 0.01014$    & $ 0.09533$  & $ 0.01125$ \\
 $0.0015$    & $ 0.07145$  & $ 0.00694$  & $ 0.07141$  & $ 0.00674$    & $ 0.07146$  & $ 0.00755$    & $ 0.07146$  & $ 0.00745$    & $ 0.07147$  & $ 0.00780$    & $ 0.07146$  & $ 0.00745$    & $ 0.07147$  & $ 0.00806$ \\
 $0.0010$    & $ 0.04761$  & $ 0.00507$  & $ 0.04759$  & $ 0.00466$    & $ 0.04762$  & $ 0.00517$    & $ 0.04762$  & $ 0.00517$    & $ 0.04762$  & $ 0.00532$    & $ 0.04762$  & $ 0.00507$    & $ 0.04763$  & $ 0.00537$ \\
 $0.0009$    & $ 0.04285$  & $ 0.00466$  & $ 0.04283$  & $ 0.00431$    & $ 0.04286$  & $ 0.00476$    & $ 0.04285$  & $ 0.00476$    & $ 0.04285$  & $ 0.00487$    & $ 0.04286$  & $ 0.00466$    & $ 0.04286$  & $ 0.00487$ \\
 $0.0008$    & $ 0.03808$  & $ 0.00431$  & $ 0.03806$  & $ 0.00395$    & $ 0.03809$  & $ 0.00431$    & $ 0.03809$  & $ 0.00436$    & $ 0.03809$  & $ 0.00441$    & $ 0.03809$  & $ 0.00426$    & $ 0.03809$  & $ 0.00441$ \\
 $0.0007$    & $ 0.03332$  & $ 0.00395$  & $ 0.03330$  & $ 0.00360$    & $ 0.03332$  & $ 0.00395$    & $ 0.03332$  & $ 0.00400$    & $ 0.03332$  & $ 0.00400$    & $ 0.03332$  & $ 0.00385$    & $ 0.03332$  & $ 0.00400$ \\
 $0.0006$    & $ 0.02855$  & $ 0.00365$  & $ 0.02854$  & $ 0.00324$    & $ 0.02856$  & $ 0.00355$    & $ 0.02855$  & $ 0.00360$    & $ 0.02856$  & $ 0.00360$    & $ 0.02856$  & $ 0.00350$    & $ 0.02856$  & $ 0.00355$ \\
 $0.0005$    & $ 0.02379$  & $ 0.00334$  & $ 0.02378$  & $ 0.00289$    & $ 0.02379$  & $ 0.00314$    & $ 0.02379$  & $ 0.00324$    & $ 0.02379$  & $ 0.00324$    & $ 0.02379$  & $ 0.00314$    & $ 0.02379$  & $ 0.00319$ \\
 $0.0004$    & $ 0.01902$  & $ 0.00299$  & $ 0.01902$  & $ 0.00258$    & $ 0.01903$  & $ 0.00284$    & $ 0.01903$  & $ 0.00289$    & $ 0.01903$  & $ 0.00289$    & $ 0.01903$  & $ 0.00279$    & $ 0.01903$  & $ 0.00284$ \\
 $0.0003$    & $ 0.01426$  & $ 0.00258$  & $ 0.01426$  & $ 0.00228$    & $ 0.01426$  & $ 0.00248$    & $ 0.01426$  & $ 0.00253$    & $ 0.01426$  & $ 0.00253$    & $ 0.01426$  & $ 0.00243$    & $ 0.01426$  & $ 0.00248$ \\
    \hline\hline
  \end{tabular}}
\end{table*}

\end{document}